\documentclass[amsmath, amssymb, pdf, floatfix]{revtex4}

\usepackage{graphicx}
\usepackage{subfigure}
\usepackage{dcolumn}
\usepackage{amsmath}
\usepackage{multirow}

\def\nn{\nonumber}
\renewcommand\deg{\mbox{${}^\circ$}}%
\newcommand\arcsec{\mbox{$^{\prime\prime}$}}%
\newcommand\arcmin{\mbox{$^{\prime}$}}%

\newcolumntype{L}{>{\centering\arraybackslash}m{1.5cm}}

\renewcommand\apj{{ApJ}}
\newcommand\apjs{{ApJS}}
\newcommand\apss{{Ap\&SS}}
\newcommand\aap{{A\&A}}
\newcommand\jgr{{J.~Geophys.~Res.}}
\newcommand\mnras{{MNRAS}}

\begin{document}


\title{A Space-based Decametric Wavelength Radio Telescope Concept}

\author{
K.~Belov$^{1}$,
A.~Branch$^{1}$,
S.~Broschart$^{1}$,
J.~Castillo-Rogez$^{1}$,
S.~Chien$^{1}$,
L.~Clare$^{1}$,
R.~Dengler$^{1}$,
J.~Gao$^{1}$,
D.~Garza$^{1}$, 
A.~Hegedus$^{1,2}$,
S.~Hernandez$^{1}$,
S.~Herzig$^{1}$,
T.~Imken$^{1}$,
H.~Kim$^{1}$, 
S.~Mandutianu$^{1}$,
A.~Romero-Wolf$^{1}$,
S.~Schaffer$^{1}$,
%
M.~Troesch$^{1}$,
%
E.J.~Wyatt$^{1}$,
%
J.~Lazio$^{1}$\\
}
\vspace{2mm}
\noindent
\affiliation{
$^{1}$Jet Propulsion Laboratory, California Institute of Technology, Pasadena, CA 91109, USA\\
%
%
%
%
%
%
%
%
%
%
%
%
$^{2}$Climate and Space Sciences and Engineering Department, University of Michigan, Ann Arbor, MI 48109-2143, USA
}

\date{\today}

\begin{abstract}
This paper reports a design study for a space-based decametric wavelength telescope.  While not a new concept, this design study focused on many of the operational aspects that would be required for an actual mission.
This design optimized the number of spacecraft to insure good visibility of approx. 80\% of the radio galaxies-- the primary science target for the mission.
A 5,000 km lunar orbit was selected to guarantee minimal gravitational perturbations from Earth and lower radio interference.
Optimal schemes for data downlink, spacecraft ranging, and power consumption were identified. 
An optimal mission duration of 1 year was chosen based on science goals, payload complexity, and other factors.
Finally, preliminary simulations showing image reconstruction were conducted to confirm viability of the mission. 
This work is intended to show the viability and science benefits of conducting multi-spacecraft networked radio astronomy missions in the next few years.  

\vspace{1pc}
\end{abstract}
\maketitle

\section{Introduction}\label{sec:intro}


The discovery by K.~Jansky of celestial radio emission occurred at a radio frequency of~20.5~MHz \cite{j35}, and discoveries at low radio frequencies ($\nu \lesssim 100$~MHz) have helped form much of the basis for modern astronomy.  Low radio frequency observations have also pioneered key technical improvements, most notably the development of aperture synthesis interferometry \cite{pp-sm46,rv46,mpp-s47,rse50,m52,r52}.  Subsequent to these early discoveries, a number of large, low radio frequency telescopes were constructed, and the field has undergone something of a recent renaissance with several telescopes being either constructed or upgraded.

A fundamental limitation to ground-based radio astronomy is the Earth's ionosphere.  Radio waves propagating through the ionosphere exhibit measurable phase changes at frequencies even well above~1~GHz, and, at frequencies below about~100~MHz, there is increasing absorption.  
The frequency at which the ionosphere becomes opaque varies both with location (due to the structure of the Earth's magnetic field) and time (e.g., diurnal effects, solar cycle), but the cutoff frequency is generally taken to be about~10~MHz. 
While there have been impressive efforts to make ground-based measurements at lower frequencies \cite[e.g.,][]{re56,c79,em87,sovw+17}, the detrimental effects of the ionosphere also are reflected in the fact that the International Telecommunications Union provides no recommendations for radio astronomical observations below 13.6~MHz~\cite{hra2004}.

At lower frequencies, space-based observations are required, and radio receivers are a common instrument on both planetary and heliospheric spacecraft \cite[e.g.,][]{wind,cassini,stereo,juno}. 
There have also been two missions dedicated to radio astronomical observations of objects beyond the solar system.  
The first Radio Astronomy Explorer (RAE-1) was in an Earth orbit and made the first measurements of the Galaxy's spectrum between~0.4 and~6.5~MHz \cite{abcsw69} while the second Radio Astronomy Explorer (RAE-2) was in a lunar orbit and observed between~25~kHz and 13~MHz \cite{akngw75}.

On the ground, there have been numerous large interferometric arrays, designed to provide higher sensitivity and imaging performance than can be realized with a single antenna, including the Giant Metrewave Radio Telescope~\cite{s90,a95}, the 74~MHz Very Large Array~\cite{vla74}, the Ukranian T-shaped Radio Telescope~\cite[UTR-2,][]{bmrsz78}, Low Frequency Array \cite{lofar}, the Long Wavelength Array~\cite{lwa1,tek+12}, and the Murchison Widefield Array~\cite{mwa}.  
There have been initial descriptions and proposals of concepts for space-based radio astronomical arrays~\cite[e.g.,][]{fhr67,wjs+88,bbd+97a,bbd+97b,op05}, notably including the Astronomical Low Frequency Array (ALFA) mission concept~\cite{alfa}, ``cubesat''-based arrays~\cite{bljsfa13}, OLFAR~\cite{Rajan2011}, DSL~\cite{Boonstra2016} and NOIRE~\cite{noire2018}.  Proofs of concept of a space-based radio interferometric array have been conducted by observing the Earth's auroral kilometric radiation with a single-element interferometer consisting of the ISEE-1 and ISEE-2 spacecraft~\cite{bgcs86} and a time-difference-of-arrival (TDOA) analysis with the Cluster spacecraft~\cite{mgc04}.

\begin{figure}[thb]
    \centering
    \includegraphics[width=0.75\columnwidth]{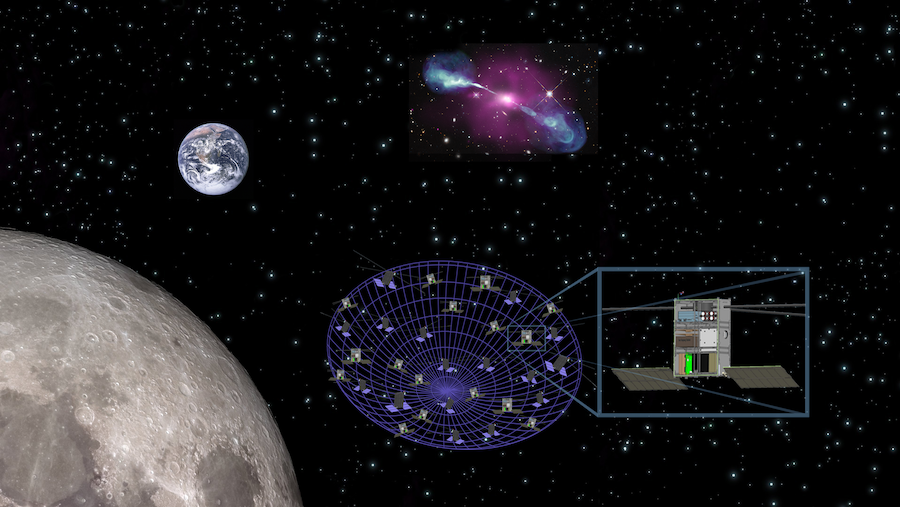}
    \caption{Artist's impression of the RELIC constellation. For clarity, the mothership is not pictured.}
    \label{RELIC-concept:fig}
\end{figure}

Recent standardization of small (``cubesat'') spacecraft components
may now make a space-based radio interferometric array feasible.
Further, the instrumentation required for detecting radio frequencies
below~15~MHz is relatively simple and amenable to a small spacecraft.
However, a recognized challenge for implementing a ``constellation''
of small spacecraft, such as would be required for a space-based
array, is the telecommunications either between spacecraft or from the
spacecraft to the Earth.  This paper reports on a concept study
designed to explore and develop solutions for implementing a
space-based array (Figure~\ref{RELIC-concept:fig}), with a primary focus is on the telecommunications
(``telecomm'') and operations.  In \S\ref{sec:science}, we summarize
a motivating science case for a space-based array and the resulting
requirements; in \S\ref{sec:spacecraft}, we summarize  strawman
spacecraft designs; in~\S\ref{sec:mission}, we present the mission
design;  in \S\ref{sec:image}, we comment briefly on aspects
related to using the transmitted data to form the desired radio
astronomical images; and, in \S\ref{sec:conclude}, we present our
conclusions.  Appendix A presents a more detailed description of the radio receiving system for the science data collection. For consistency with the concept described in
\cite{kiss}, we retain its name of \hbox{RELIC}.

\section{Science Definition and Mission Design Requirements}\label{sec:science}

The potential science cases for a space-based array have been well described in previous discussions of such a concept \cite[e.g.,][]{fhr67,wjs+88,alfa,op05,bljsfa13,Rajan2011,Boonstra2016, kw90, swgb00}.  

The science cases cover a large range of topics including radio
emission from energetic particles in the inner heliosphere,
magnetospheric emissions from solar system planets with planetary-scale
magnetic fields, and, potentially in the future, studying
magnetospheric emissions from extrasolar planets.
While we considered the above mentioned cases, we choose to focus on a specific science case, namely the imaging of radio galaxies.
This choice is motivated in part by the identification of this science case as one that forms a key part of a multi-wavelength study of physical processes in active galaxies \cite{kiss} and in part because this science case produces requirements that are more broadly applicable to a number of other science cases.

In brief, the combination of synchrotron and inverse Compton emission,
the former measured at radio wavelengths and the latter measured at
X-ray wavelengths, provide information on the plasma and magnetic
field properties in regions where particles are being accelerated or
as they stream away from those regions.  Ground-based measurements can
be used to measure a portion of the synchrotron spectrum
\cite[e.g.,][]{h05,ichm17}, but access to the radio spectrum
at~10~MHz or lower would reduce uncertainties substantially by
limiting the extent to which the electron energy distribution has to
be extrapolated in energy.  In order to have some overlap with
ground-based observations, we specify a frequency range below~30~MHz.

\begin{figure}[thb]
  \begin{minipage}[t]{0.45\textwidth}
    \centering
    \includegraphics[width=\columnwidth]{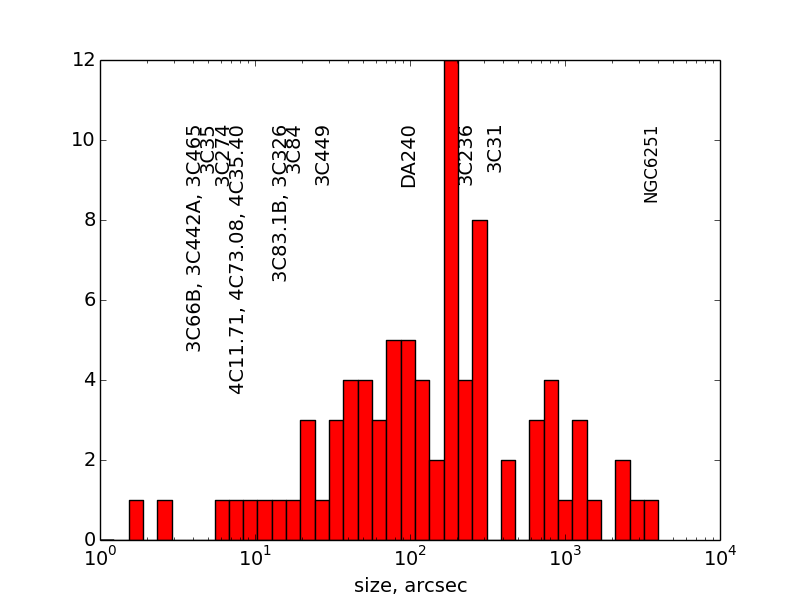}
    \caption{DRAGN size distribution from the 3CR~\cite{Laing1983}.}
    \label{dragn_size:fig} 
  \end{minipage}
  \begin{minipage}[t]{0.45\textwidth}
    \centering
    \includegraphics[width=\columnwidth]{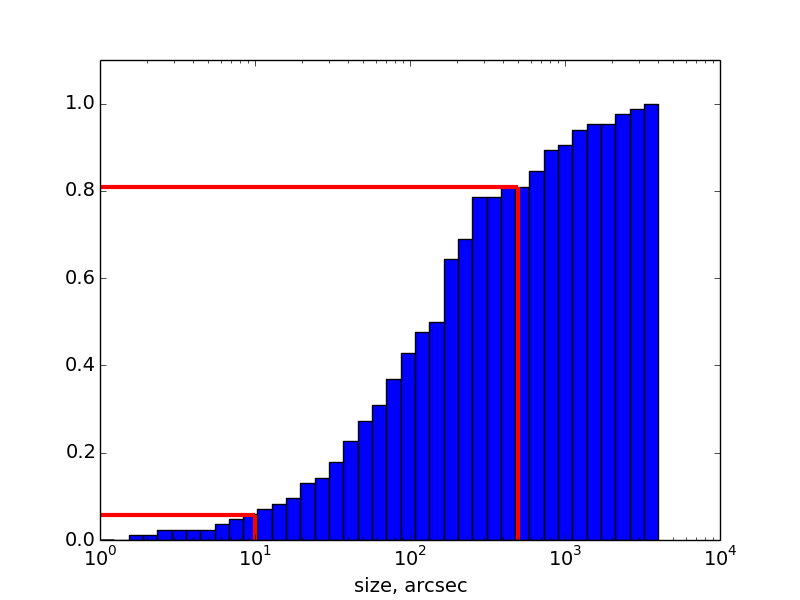}
    \caption{Cumulative distribution of DRAGN angular sizes.  The red lines demarcate the range encompassing approximately 80\% of the sources.}
    \label{dragn_size_cumulative:fig} 
  \end{minipage}
\end{figure}

In order to set high-level science requirements for the RELIC concept,
we selected the nearest 85 double radio sources associated with a
galactic nuclei (DRAGNs) from the 3CR sample \cite{Laing1983}.
Figures~\ref{dragn_size:fig} and~\ref{dragn_size_cumulative:fig} summarize the range of angular scales
observed in DRAGNs, with the mean size being approximately 100\arcsec\
and 80\% of the sources being in the range 10\arcsec\ to~500\arcsec.
In addition to the intrinsic sizes of DRAGNs, at the frequencies of
interest, significant angular broadening due to density fluctuations
in the interstellar medium occurs \cite{Cordes1990}.  This propagation
effect limits to approximately 10\arcsec\ 
the finest angular resolution that is required.

\begin{figure}
  \begin{minipage}[t]{0.45\textwidth}
    \centering
      \includegraphics[width=\columnwidth]{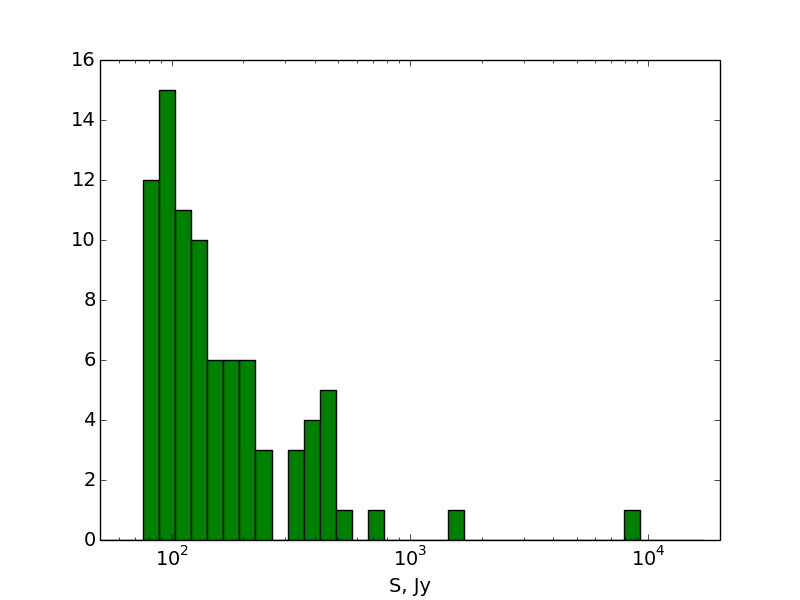} 
      \caption{Flux densities of our double radio sources associated with a galactic nuclei (DRAGNs) sample, as selected from the 3CR survey and scaled to~10~MHz.  The truncation at low flux densities represents the completeness limit of the parent catalog from which our sample is drawn.}
      \label{fig:dragn_flux}
  \end{minipage}
  \begin{minipage}[t]{0.45\textwidth}
    \centering
      \includegraphics[width=\columnwidth]{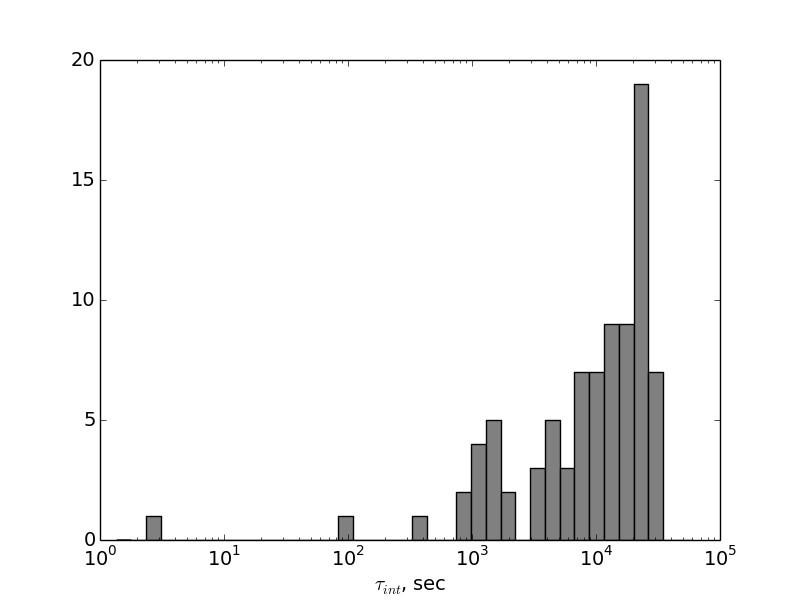}
      \caption{DRAGN integration time at 10~MHz. The calculation is based on the flux densities shown in Figure~\ref{fig:dragn_flux}}
      \label{DRAGN_integration_time10MHz:fig} 
   \end{minipage} 
\end{figure}

Figure~\ref{fig:dragn_flux} shows the distribution of flux densities at~10~MHz for our sample.
In order to estimate these flux densities, we scaled the measured flux densities from their measured values at~178~MHz assuming a spectrum of $S \propto \nu^\alpha$, with $\alpha = -0.7$.
This scaling assumes that there is no free-free absorption due to a foreground thermal population of electrons.
Some surveys show significant variations of the scaling for different sources \cite{Braude:2002}.
There is some evidence of the Cygnus A hotspot spectrum turn over at lower frequencies~\cite{McKean2016}, but generally, the radio emission of interest from these DRAGNs is from well outside the main stellar disk of the host galaxy and they are at high Galactic latitudes, meaning that this assumption should be valid for the vast majority, if not all, of our sample.

Table~\ref{tab:sdr} summarizes these science requirements for the
RELIC mission concept.


\begin{table}[h]
\caption{Science Driven Requirements}
\label{tab:sdr}
\centering
\begin{tabular}{|c|c|c|} \hline
{\bf Requirement}                         & {\bf Science}         & {\bf Implied Mission Requirements} \\
\hline
{\bf Angular resolution} & 10\arcsec\protect\footnote{Limited by interstellar broadening} & 620 km max S/C separation \\
\hline
{\bf Largest angular scale} & 30\arcmin\protect\footnote{Angular size of a large radio galaxy} & \begin{tabular}{@{}c@{}}12 km min S/C separation\\Non-repeating baselines\\32 S/C minimum\end{tabular} \\
\hline
{\bf Frequency bandwidth} & \begin{tabular}{@{}c@{}}0.1-10 MHz + ~ 30~MHz\\Best science $<$~10~MHz\\Overlap with ground: 30~MHz\end{tabular} & Data rate \\
\hline
{\bf Image S/N ratio} & 10 & Min source obs. time 3 hours \\
\hline
\end{tabular}
\end{table}

Table~\ref{tab:mdc} summarizes the high-level mission requirements
that are derived from the science requirements (Table~\ref{tab:sdr}).

\begin{table}[h]
\caption{Requirements driven by mission design considerations.}
\label{tab:mdc}
\centering
\begin{tabular}{|c|c|c|} \hline
{\bf Parameter}                         & {\bf Value}         & {\bf Rationale} \\
\hline
{\bf Frequency resolution} & 16~kHz & See sec.~\ref{pf:subsec}\\
\hline
{\bf Spectral windows } & $>$~64 & See sec.~\ref{fft:subsec}\\
\hline
{\bf Dynamic range} & $10^3$--$10^4$ & S/N ratio, see table~\ref{tab:sdr} and sec.~\ref{rdd:sec}\\
\hline
{\bf Polarizations} & 2 & See below in this section  \\
\hline
{\bf Sessions per source} & 1--3~hour \protect\footnote{Or multiple sessions totaling $\sim$3~hours} & Can be split into mult. sessions \\
\hline
\end{tabular}
\end{table}

A fundamental telescope architecture question is whether it should be monolithic (i.e., completely filled aperture) or segmented; we consider an interferometer to be an extreme example of a segmented telescope.
Simple optics considerations show that, in order to obtain the required angular resolution at the relevant frequencies, an effective aperture exceeding 100~km in diameter is required.
Consequently, we henceforth adopt an interferometric design for the telescope.

The maximum angular resolution of the array is limited to $\sim$10\arcsec~by the interstellar broadening \cite{Cordes1990}. 
At 10 MHz this corresponds to 620~km longest projected baseline, which is already presents significant communication challenges, limiting the available data bandwidth.
The dilution of the angular resolution can also be due to uncertainty in the asset location positing and due to the clock drift, which imposes corresponding hardware requirements.
In order to provide the best source visibility, the minimum spacecraft (S/C) separation should be less than 12~km, as the largest radio galaxies are about 30\arcmin~in size.  
The baseline mission architecture is one \emph{mothership} with~32 \emph{daughterships} delivered to the initial orbits by a single launch vehicle.  
Given a target flux density, a target S/N ratio, and a notional integration time, see Figures \ref{fig:dragn_flux} and~\ref{DRAGN_integration_time10MHz:fig}, an array composed of 32 dual-polarization antennas is sufficient to detect a typical source.

Each S/C should carry a dual-polarization dipole, i.e., be capable of receiving both polarization, in order to provide an additional $\sqrt{2}$ improvement in signal-to-noise ratio and to provide redundancy for each \hbox{S/C}.
In addition, signal polarization is a powerful means of rejecting terrestrial interference.  Not considered in this trade study was a cost-benefit analysis of each S/C carrying a so-called tripole system in order to capture the full three-dimensional information about the electric field~\cite[e.g.,][]{KleinWolt2012}, but one immediately obvious cost of carrying a tripole antenna system on each S/C would be an increased data rate requirement.
One design solution for a tripole antenna is described in details in \cite{Chen2018}

Finally, the mission should allow for the S/C pointing to keep the sources in the FOV during a session long enough to establish an interferometric image. 
The orbits are discussed in details in the section~\ref{sec:mission.orbits} below.
The ability to sample signal up to 30~MHz is desirable to overlap with the ground observations. 
This requirement dictates the sampling rate of 60~MHz and a broadband antenna, the exact design of which will be chosen later.
Anticipating future results as discussed below, we consider the observation bandwidth to be 16~kHz.
Each daughtership performs very limited data processing, downlinking the data to the mothership for further processing and further downlinking to the ground. 
This architecture is a compromise between the limited bandwidth for the data transfer to the mothership, limited power available to the daughtership and limited or non-existent daughtership to daughtership intercom.


\section{Mission Architecture}\label{sec:mission}


\subsection{Design Optimization Using Operations Analysis}\label{sec:mission.optimization}

One of the challenges in space mission design is correctly accounting for a large number of design dimensions that may interact in subtle and hard to predict ways.  We address this difficulty by adopting an operations-based approach to evaluating mission designs.
We in effect partially simulate the missions, applying any and all operations constraints we can to derive results as realistic as possible.
We characterize the science measurements possible and use these as a proxy for mission return.
By performing these simulations and calculations, we aim to estimate mission return and therefore enable devoting resources to the most promising early mission designs.

In this section we discuss our implementation and results, focusing on a data throughput analysis to estimate mission lifetime to fulfill its main scientific goals.

\subsubsection{Data Throughput Analysis Using Operations Planning}\label{sec:mission.optimization.throughput}

The data throughput analysis is performed using the ASPEN/CASPER planning system~\cite{Chien2000_6, Chien2000_4}.
ASPEN is a timeline-based scheduling framework that allows for operations, \hbox{S/C}, science, and other constraints to be incorporated in an automated scheduling environment.
The automatic scheduling algorithms generate a proposed mission operations schedule following a set of constraints. 
The generated mission plans may then be evaluated for various metrics
such as science data utility and remaining resource margins.

The planner leverages a common core of action/state models. 
The actions available in the common model span the entire constellation: some are executed only on individual daughterships, some only on the mothership, and others require joint simultaneous action by multiple spacecraft. 
The modeled actions include repointing the daughterships, science data acquistion, crosslinking data from a daughtership to the mothership, downlinking data from the mothership to Earth, downlinking data directly from a daughtership to Earth, and placeholders for intermittently required engineering activities. 
These actions make use of various modeled states and resources including the visibility of each science target, the interferometric baseline utility of each observation window, the number of receivers on the mothership, the visibility of Deep Space Network ground stations, the bandwidth of each communication link, power generation rate, and remaining battery reserves.

The CASPER automated operations planning system then uses the combined core model and design constraints to generate a proposed operations plan. 
CASPER starts from an empty mission plan and iteratively optimizes it by adding or removing actions to improve a declared utility function. 
The utility function is directly related to the calculated science utility of the data received at Earth, and strongly inversely related to any mission constraint violations. 
This approach guides the planner to add observation, crosslink, and
downlink activities while respecting the design limits on view
periods, storage space, bandwidth, power, and other similar constraints.
The calculated utility of the science data is related to the total observation integration time and how well the selected interferometry baselines cover the space of distances and angles needed to characterize the structure of each radio astronomy target. 
The final output operations schedule from the planner includes concrete timed actions for each of the constellation craft to execute.

Critical to all of this operations planning is the geometric aspect of the problem.  
For all of these geometric analyses, we use the SPICE package~\cite{Acton1996}.
These analyses include daughtership position and science target position for array analysis, daughtership and mothership position for crosslink calculations, and mothership and ground station position and downlink calculation.

The operations schedule can then be evaluated versus specified metrics.
Primary metrics are the uv-plane coverage, integration time, and target coverage. 
Additional metrics such as excess unused capacity on some resources (e.g., unused power or bandwidth) are also reported to help inform which parts of a design may be over-engineered and which are the bottlenecks during actual operations.
For the data throughput analysis, the duration of crosslink, downlink, and observation activities are of interest.

\begin{figure}[thb]
 \includegraphics[width=\columnwidth]{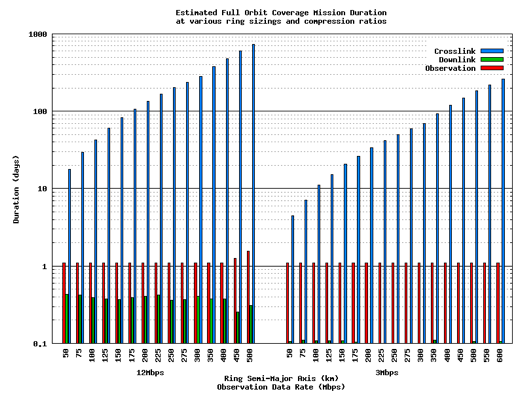}
 \caption{
Intervals for relevant activities per day of observation as a function of crosslink data rate and
maximum separation between daughterships, with the different color
bars indicating the different activities---crosslink (blue), downlink
(green), and observation (red).}
 \label{duration_for_science:fig}
\end{figure}

Figure~\ref{duration_for_science:fig} shows the amount of time that would be required for science data acquisition, crosslink from daughtership to mothership, and downlink from mothership to Earth, all determined using the approach described above.
It is clear that crosslinking would be the most costly action in terms of mission duration, requiring roughly 10 mission days per observation day. 
The crosslink time also would increase significantly with the separation of the daughterships, while the observation time and the downlink time would remain constant.
As the daughterships move further away from the mothership, assumed to be at the center of the array, the available bandwidth for crosslinking decreases.
One technique to offset the decrease is mentioned below. 

\subsection{Orbit Selection and Modeling}\label{sec:mission.orbits}

\begin{figure}[thb]
 \includegraphics[width=\columnwidth]{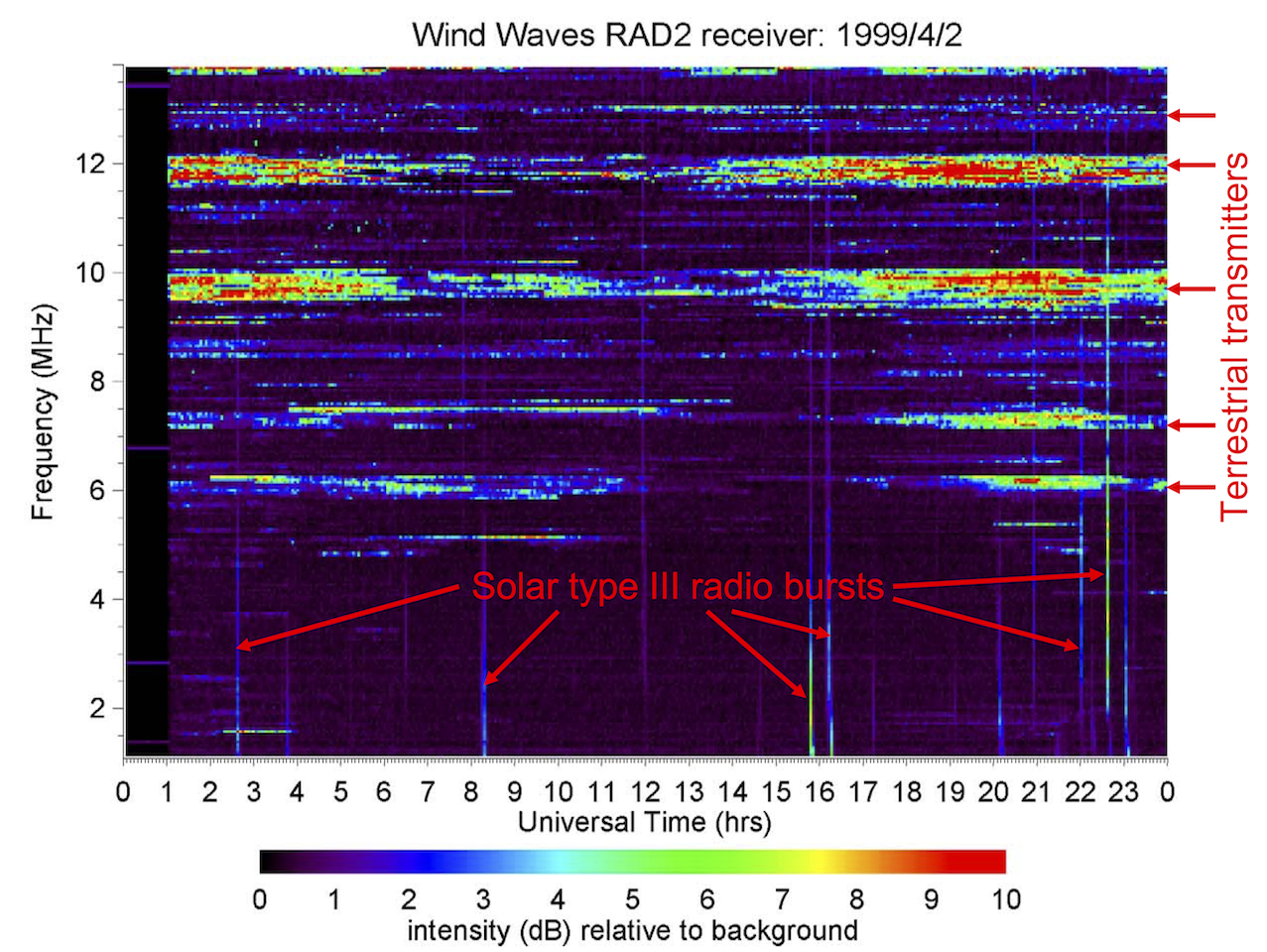}
 \caption{Dynamic spectrum from the Wind/WAVES instrument, when the Wind spacecraft was approximately at one lunar distance on April 2, 1999. \hbox{NASA/GSFC}, R.~MacDowall..}
 \label{Wind_Waves_RAD2_19990402:fig}
\end{figure}

The baseline orbit design is a mothership orbiting in a circular orbit with an altitude of~$5000$~km above the Moon and 32 daughterships orbiting around the mothership in a cluster.
The number of dautherships was chosen to serve as a ``baseline'' to obtain an adequate uv-plane coverage from a preliminary analysis with several major assumptions made about the hardware.
Once the hardware design is finalized during the mission proposal study, a more accurate modeling will be possible. 
The mothership orbit was chosen for science purposes to be far enough from the Earth to avoid interference, as well as minimize gravitational perturbations from Earth.
Figure~\ref{Wind_Waves_RAD2_19990402:fig} shows a dynamic spectrum from the Wind/WAVES instrument, when the Wind spacecraft was approximately at one lunar distance on April 2, 1999.
While there is some ionospheric breakout, below 5 MHz, there is essentially no terrestrial RFI.  Even above 5 MHz, there are still relatively clean bands at least up to 10 MHz.

\begin{figure}[h]
\centering
 \subfigure[Inertial, Fixed Frame]{%
  \includegraphics[width=.45\textwidth]{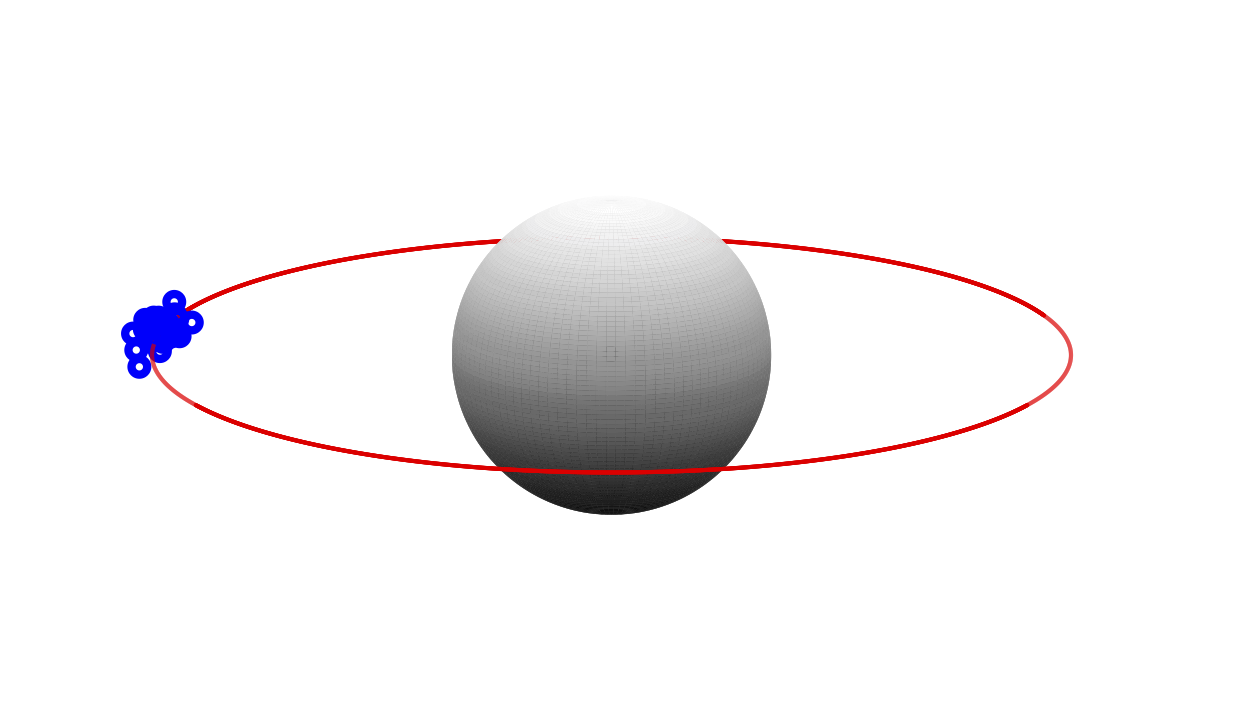}
		}
 \subfigure[Relative, Rotating Mothership Frame]{%
  \includegraphics[width=.45\textwidth]{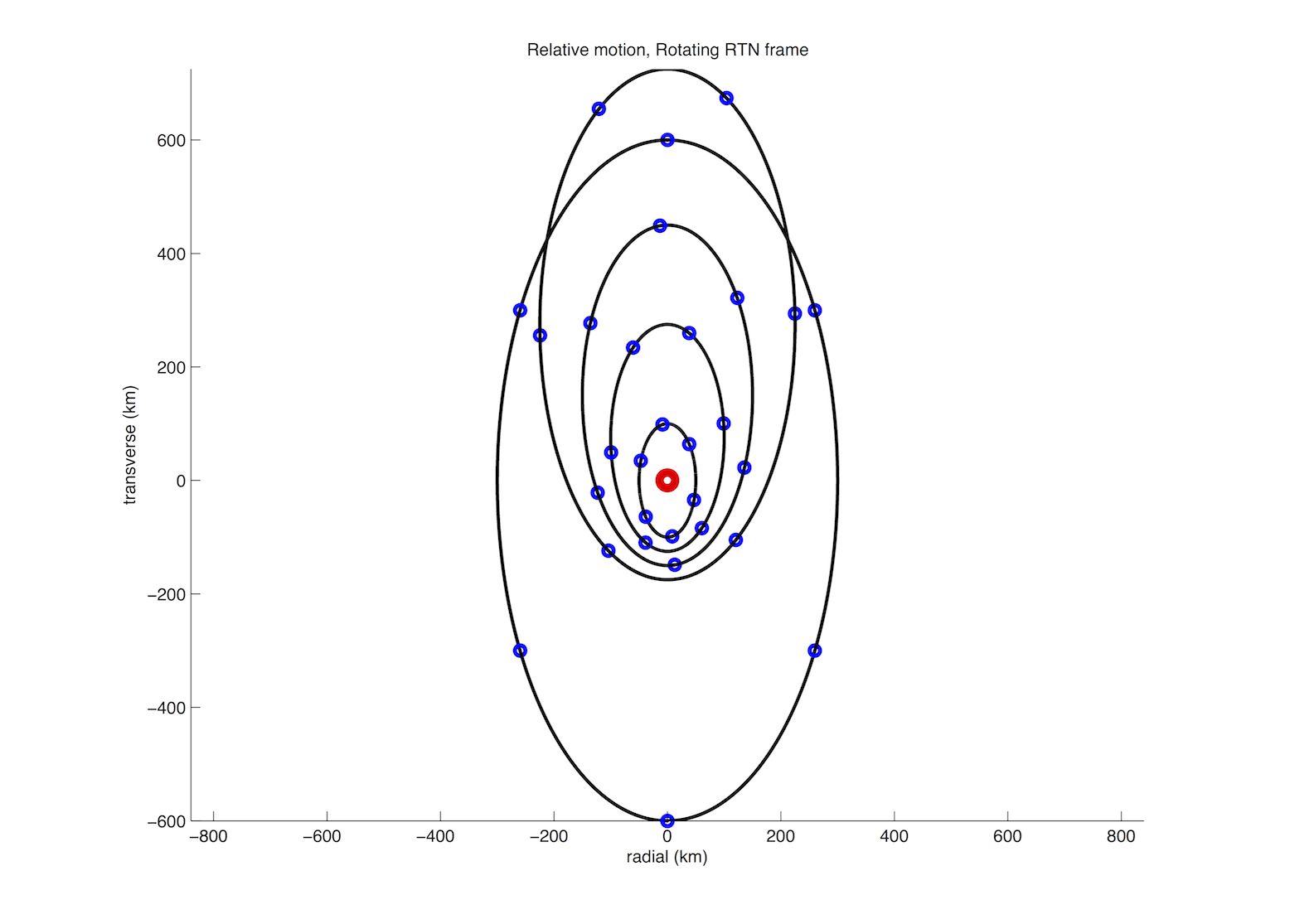}
		}
 \caption{Orbit design for mothership and daughterships in inertial
and relative frame.  The orbits of the daughterships are termed
``rings.''}
 \label{fig:initial_orbit_design} 
\end{figure}

Due to perturbations from Earth and Sun, the chosen orbits vary in inclination over the course of time, from -30 deg. To +30 approximately.
Although not required by the mission, the brief periods of RFI shielding from Earth are an added benefit.
The maximum allowed baseline between any two daughterships for science is approximately 600~km; therefore, the entire cluster needs to remain relatively close (Figure~\ref{fig:initial_orbit_design}a).
For this reason, the orbits of all the daughterships are designed to have the same period ($\approx 8.8$~hr), with slightly varying eccentricity and inclination.
In a relative, rotating frame fixed at the mothership, the equal period orbits are accomplished by 2$\times$1 ellipses of varying sizes and centers, mimicking a ``gear-like'' movement, which allows for optimum uv-plane coverage (Figure~\ref{fig:initial_orbit_design}b).
We henceforth term these ellipses on which the daughterships would be orbiting as ``rings.''

As the orbital period is constant, after one full revolution around
the Moon, the daughterships all would return to the initial relative
configuration, i.e., same location in
Figure~\ref{fig:initial_orbit_design}b.  Most of the science and data
downlink could be accomplished within a few orbits around the Moon for
a specific configuration.  Because the structure of the target radio
galaxies is not expected to change significantly on the time scale for
this mission, it is possible to improve the aperture plane coverage by
reconfiguring each ring to change its size.  This technique is
conceptually similar to the practice used in early efforts to
reconstruct the brightness distribution of radio galaxies
\citep{m53,b57,rh60} and continues to be used to date, e.g., combining
data from multiple configurations of the Very Large Array.  The main
constraint is to maintain the same orbit period for all daughterships,
which can be accomplished only via a maneuver perpendicular to the
velocity direction.  For each ring, the maneuver for reconfiguration
would happen at the same location in the relative orbit, ensuring that
all daughterships reconfigure to the same ring.  It takes about one
orbit revolution to reconfigure all daughterships on one ring.
Reconfiguration should happen sequentially from largest to smallest
ring in order to minimize the risk of collisions.  The daughterships
could reconfigure as many times as needed, within a given $\Delta V$
budget.

\begin{figure}[h]
 \centering
 \subfigure[Initial Configuration]{%
	\includegraphics[width=.8\textwidth]{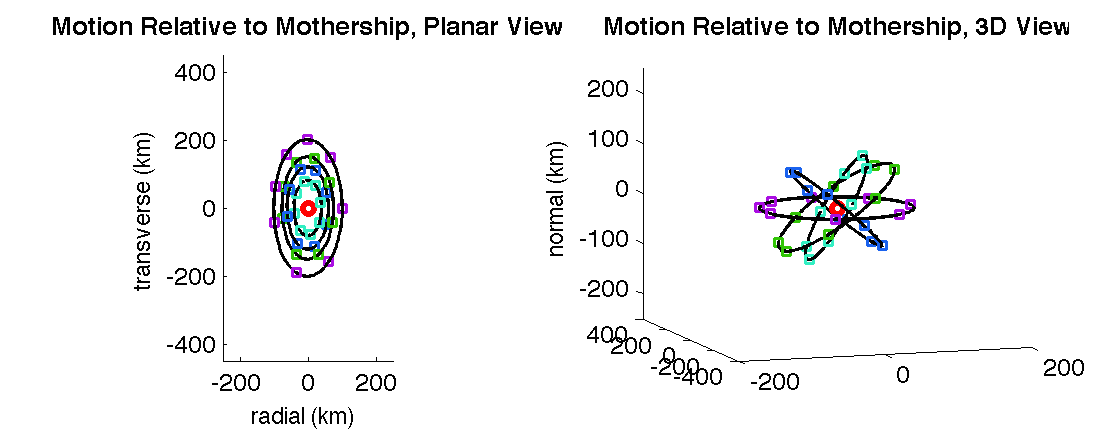}
		}
 \subfigure[Final Configuration]{%
	\includegraphics[width=.8\textwidth]{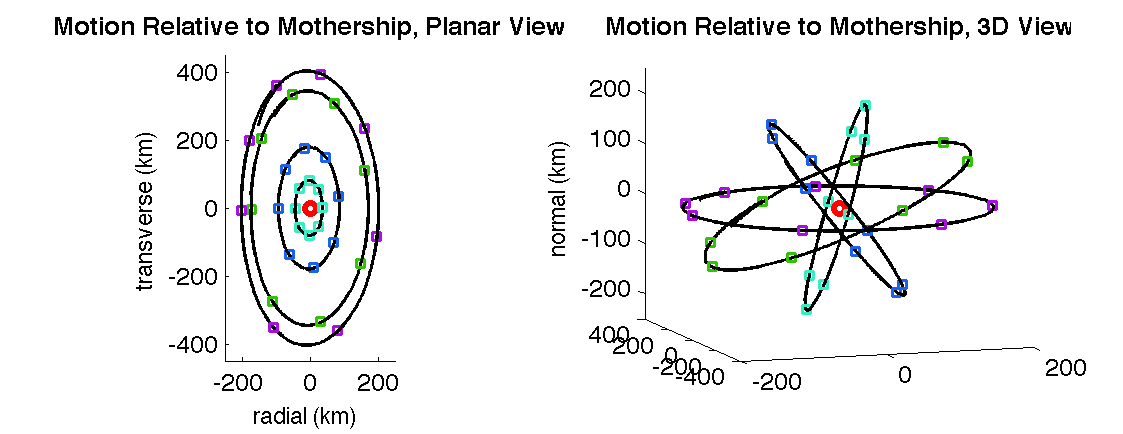}
		}
 \caption{Reconfiguration of the daughterships from their initial
configuration to the final configuration.  In all panels, the
configurations are shown in a relative, rotating frame fixed at the
mothership, which is represented by the red dot at the origin.}
 \label{fig:initFinOrbit}
\end{figure}

The final design chosen is four rings, with~8 daughterships in each ring. 
Each daughtership would deploy from the mothership sequentially to its
initial configuration (Figure~\ref{fig:initFinOrbit}[a]; Table~\ref{tab:config}).
The mothership is at the center of the formation.
The initial configuration of the rings has been designed to cost 20~m~s${}^{-1}$ for each daughtership.
After all the science data has been gathered and downlinked to Earth
within a specific configuration, the daughterships would reconfigure, applying 1~m~s${}^{-1}$ perpendicular to their velocity, which allows for 5~km increase in ring size.

Such a ring reconfiguration could be obtained if each daughtership
carries a 1~N thruster capable of generating a specific impulse of at
least $I_{\mathrm{s}} = 200\,\mathrm{s}$ (\S\ref{sec:spacecraft}).
A total of~20 reconfigurations (equivalent to~20~m~s${}^{-1}$ of propellant) could applied, ending in the final configuration shown in Figure~\ref{fig:initFinOrbit}(b) with parameters in Table~\ref{tab:config}.

\begin{table}[h] \caption{Initial and Final Orbit Sizes with Respect to the Mothership of Each Ring } \label{tab:config}
\centering
\begin{tabular}{|c|c||c|c||c|c|} \hline
 & &\multicolumn{2}{|c||}{Initial Configuration} & \multicolumn{2}{|c|}{Final Configuration} \\ \hline
Ring & Color & Semi-Major Axis & Inclination & Semi-Major Axis & Inclination \\
 & (Fig.~\ref{fig:initFinOrbit}) & (km) & (deg)  & (km) & (deg) \\ \hline \hline
1 & purple & 200 & 0.0 & 400 & 0.0 \\
2 & green & 160 & 25.5 & 300& 18.5\\
3 & blue & 140 & 150.0 & 200 & 138.0 \\
4 & cyan & 100 & 45.0 & 100 & 63.5 \\ \hline
\end{tabular}
\end{table}


\subsection{Coverage and Baseline Computations}

The formation design is driven by adequate coverage of collection targets, with coverage defined by the diversity of baselines formed by individual daughtership pairs. 
A single baseline is the projection of the relative position vector
from one daughtership to another, into the plane perpendicular to the direction to a target.
Baselines are computed beginning with the relative position vector
from daughtership~$i$ to daughtership~$j$, $\vec{\rho}_{ij}$, and the unit vector to a target, $\hat{e}_\star$. 
Both vectors are assumed to be in a common inertial frame, EME2000 for this study. 
First, a new frame for projecting the position vectors is formed using the target vector and the z-axis unit vector in the EME2000 frame:
\begin{align}
\nn  \hat{e}_z & = \hat{e}_\star \\
\nn \hat{e}_y & = \hat{k}   \times \hat{e}_z \\
\nn  \hat{e}_x & = \hat{e}_y \times \hat{e}_z
\end{align}
where $\hat{e}_x$, $\hat{e}_y$, and $\hat{e}_z$ are the unit vectors defining the target frame; and $\hat{k}$ is the unit vector in the EME2000 $z$ direction.
For targets near $90^\circ$ declination, the EME2000 $x$ direction can be used in place of the z direction.
With this target frame constructed each position vector $\vec{\rho}_{ij}$ is projected into the $xy$-plane of the target frame to yield $(r,\theta)$ pairs describing the baseline:
\begin{align}
\nn  r      & = \sqrt{b_x^2 + b_y^2} \\
\nn  \theta & = \tan^{-1}\left(b_y/b_x\right)
\end{align}
where $b_x$ and $b_y$ are the components of $\vec{\rho}_{ij}$ in the $x$ and $y$ directions of the target frame, respectively ($b_x  = \vec{\rho}_{ij} \cdot \hat{e}_x$ and
 $b_y = \vec{\rho}_{ij} \cdot \hat{e}_y$).
 We operate with $(r,\theta)$ terms instead of usual $u,v$ components for convenience given the geometry of the problem.
 When computing baselines it does not make a difference if the baseline is measured from daughtership~$i$ to~$j$, or vice-versa.
 Because of this a baseline of $(r,\theta)$ counts the same as one in the opposite direction, $(r,\theta + 180^\circ)$, and a formation of~$n$ daughterships will have $n (n - 1) / 2$ unique baselines at a given sample time.
 For the purpose of computing coverage the $(r, \theta)$ space is divided up into $N$ bins in the $r$ direction and $M$ bins spanning $180^\circ$ in the $\theta$ direction, and credit is taken for each bin covered by a baseline.
 As the formation orbits the Moon and formation geometry changes, new baseline measurements are taken, and the overall coverage consists of the different $(r, \theta)$ bins which have been collected.
 If $N_{\mathrm{collect}}$ is the number of bins collected over some period of time then a numerical coverage score, $J$, can be computed from $J = N_{\mathrm{collect}} / (N M)$.

\begin{figure}[h]
\centering
   \subfigure[Initial Baselines]{%
	\includegraphics[width=.49\textwidth]{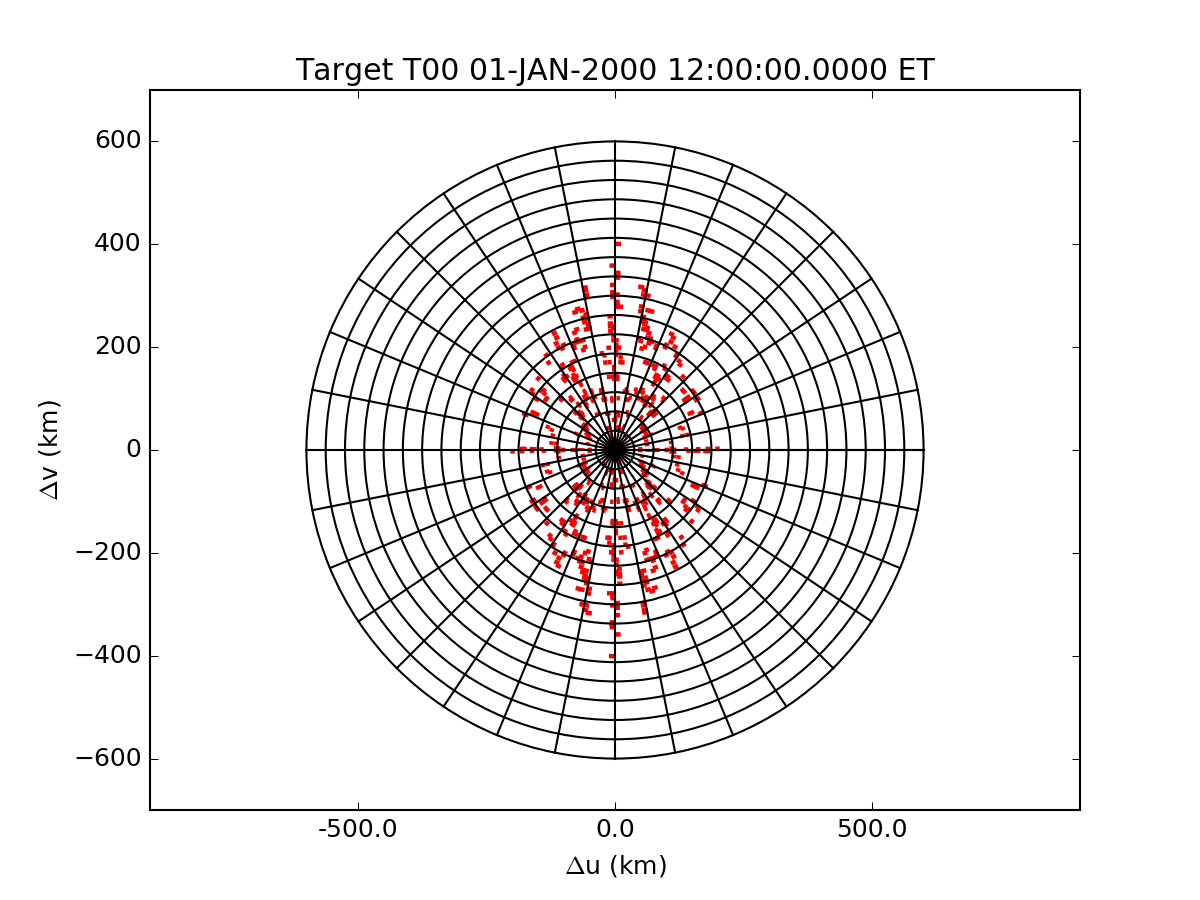}
		}
   \subfigure[Cumulative After One Orbit]{%
	\includegraphics[width=.49\textwidth]{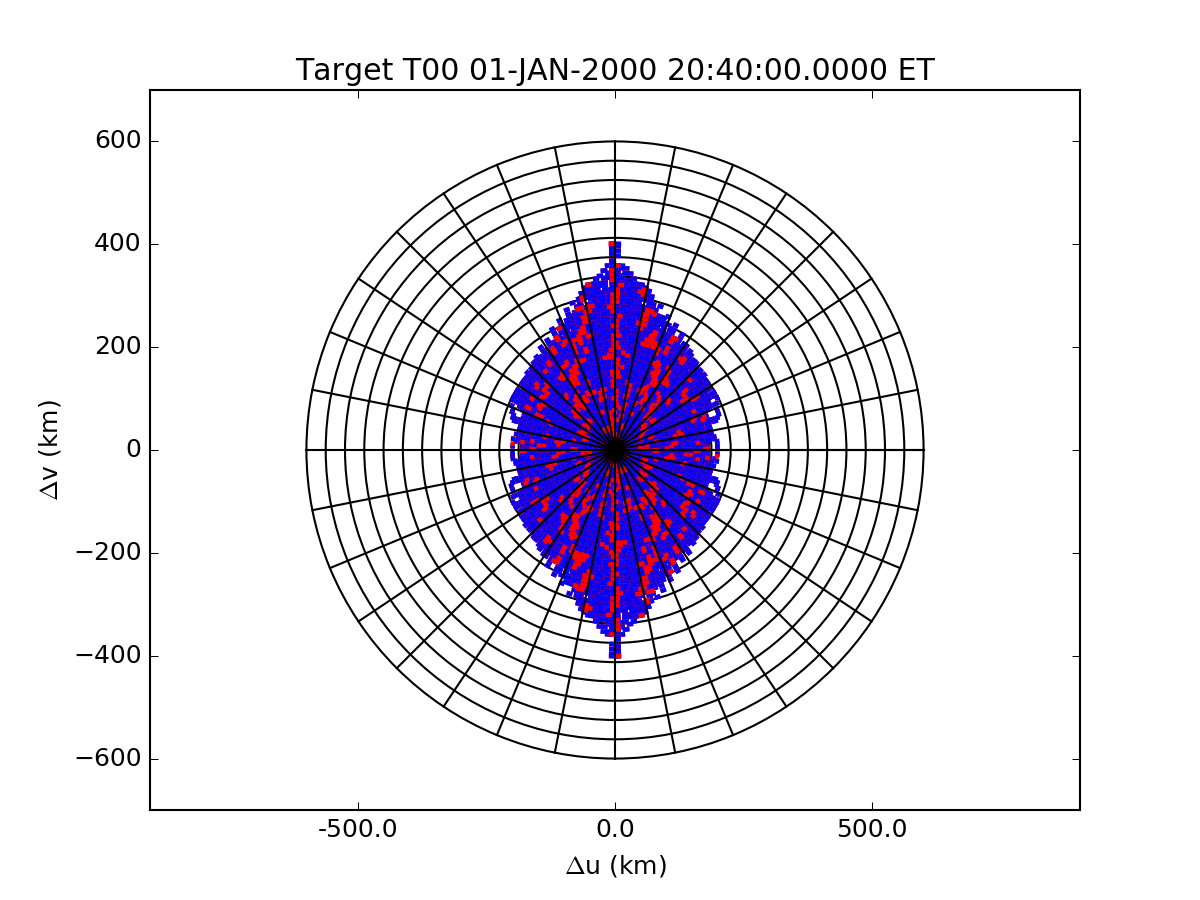}
		}
   \caption{%
The uv-plane coverage for the initial configuration of daughterships.  In
this example, the target radio galaxy is at considered to be at a
right ascension of $0^\circ$ and declination of $0^\circ$; similar
coverage diagrams apply for sources at comparable declinations.
A total of~128 bins in the radial direction and~128 bins to cover $180^\circ$ in $\theta$, with gridlines every 8 bins.
This number of bins seems to be optimal for up to~30\arcmin~size of the radio galaxy.
Red bins indicate the instantaneous baselines at the time given while blue bins show bins already collected, with samples taken every 10~minutes.
The left panel shows the initial set of baselines collected while the
right panel illustrates the cumulative baselines acquired after one orbit period.}
 \label{fig:tgt00Baselines}
\end{figure}

\begin{figure}[h]
\centering
   \subfigure[Initial Baselines]{%
	\includegraphics[width=.49\textwidth]{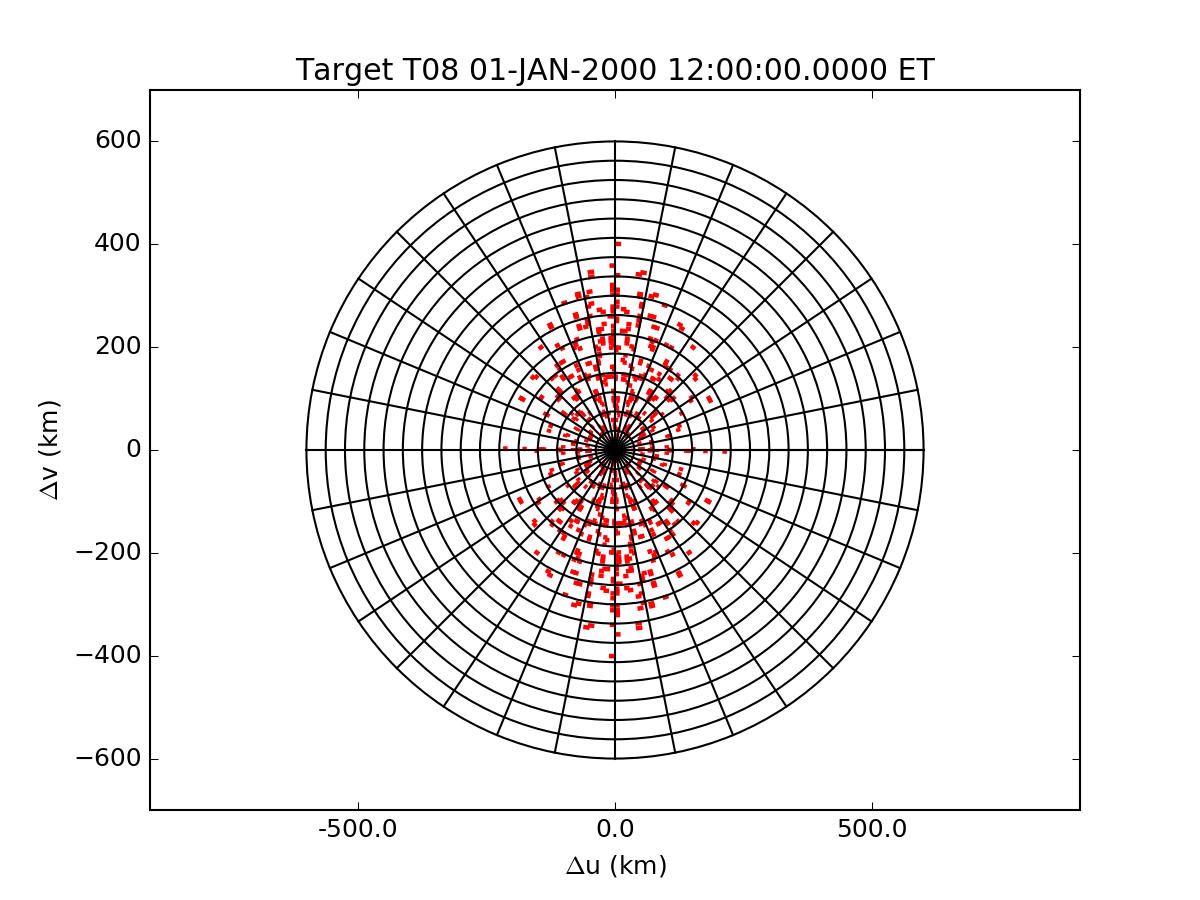}
		}
   \subfigure[Cumulative After One Orbit]{%
	\includegraphics[width=.49\textwidth]{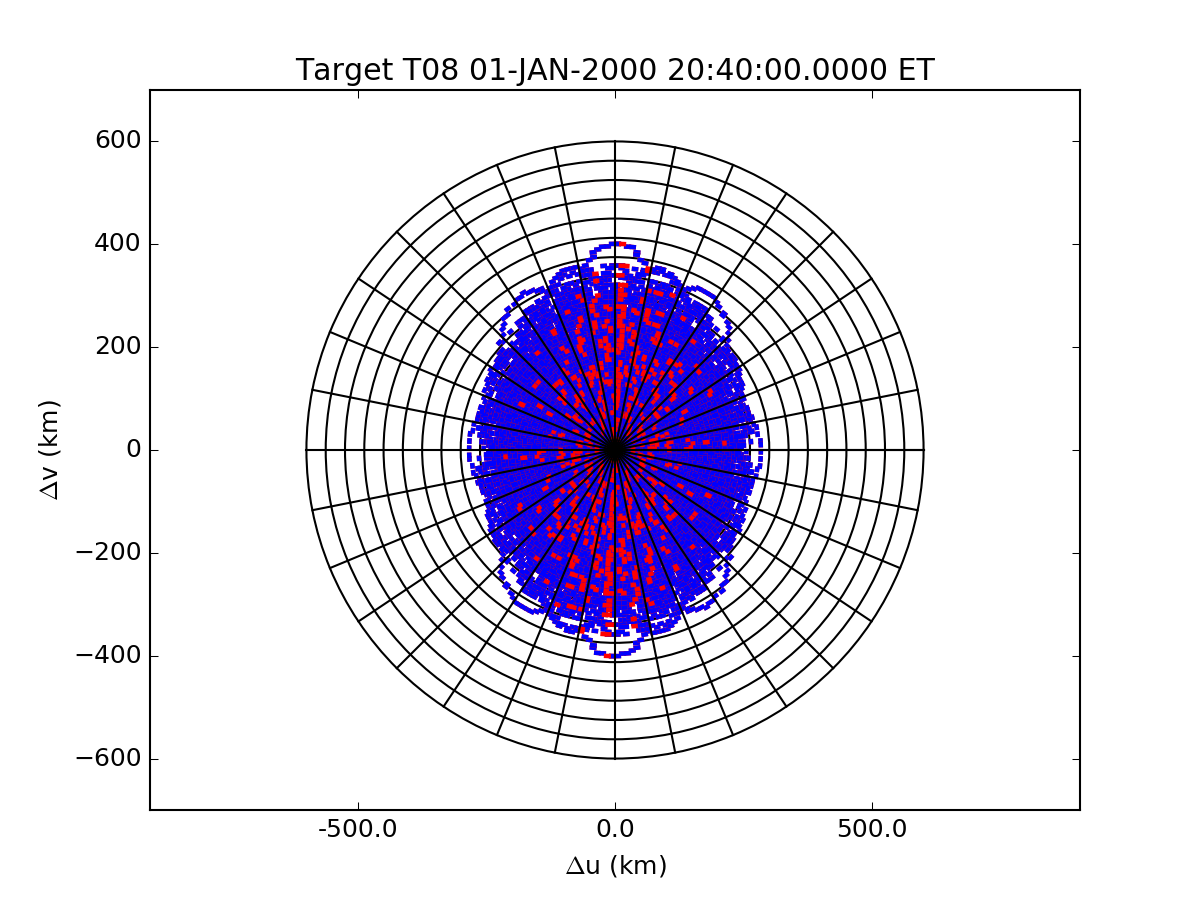}
		}
   \caption{%
The uv-plane coverage for the initial configuration of daughterships.  In
this example, the target radio galaxy is at considered to be at a
right ascension of~$0^\circ$ and declination of~$45^\circ$; similar
coverage diagrams apply for sources at comparable declinations.
A total of~128 bins in the radial direction and~128 bins to cover $180^\circ$ in $\theta$, with gridlines every 8 bins.
Red bins indicate the instantaneous baselines at the time given while blue bins show bins already collected, with samples taken every 10~minutes.
The left panel shows the initial set of baselines collected while the
right panel illustrates the cumulative baselines acquired after one orbit period.}
\label{fig:tgt01Baselines}
\end{figure}

Figures~\ref{fig:tgt00Baselines} and~\ref{fig:tgt01Baselines} show
uv-plane coverage for the formation in the initial configuration in
Table \ref{tab:config}, against two different targets in different directions.
The different coverages obtained reflect the assumed target source declinations, and the resulting differences in collection geometry.
As the mothership orbit lies in the EME2000 $xy$-plane, the cumulative
baseline pattern will be flatter against targets with low
declinations; the pattern will be more round against targets with high declinations.
What thickness exists in the coverage pattern against low-declination
targets is due to the out-of-plane components of the outer rings.
Once the final configuration is achieved, the coverage pattern will cover a greater area.

\subsection{Communications and Data Downlink}\label{sec:mission.comm}
\subsubsection{RELIC Communications Architecture Definition}\label{sec:mission.comm.arch}

\begin{figure}[thb]
\begin{minipage}[t]{0.3\linewidth}
\centering
\includegraphics[width=\columnwidth]{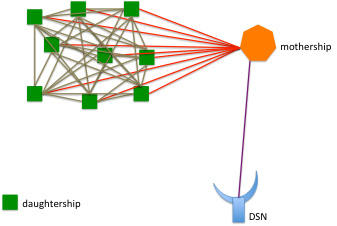}
\end{minipage}
\begin{minipage}[t]{0.3\linewidth}
\centering
\includegraphics[width=\columnwidth]{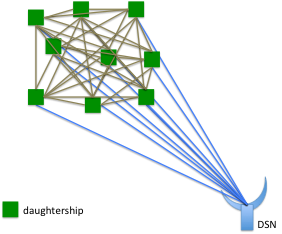} 
\end{minipage}
\begin{minipage}[t]{0.3\linewidth}
\centering
\includegraphics[width=\columnwidth]{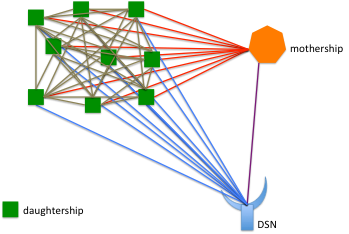} 
\end{minipage}
\caption{Potential constellation communications topologies for a
space-based radio astronomy array.  
For clarity, the mothership (orange) is shown spatial offset from the center of the constellation, but, in the actual design, it would be located near the center in order to minimize the communication range.}
\label{commtopology:fig}
\end{figure}

Figure~\ref{commtopology:fig} illustrates a few examples of the topologies derived from our communications trade analysis.
The daughtership-to-daughtership mesh is primarily used for ranging, discussed further below. 
Direct-to-Earth (DTE) daughtership communications were considered for
a mission concept that would not include a mothership, but the
communication demands on a small spacecraft, for the required data
rates, were judged infeasible.
In general, the communications architecture and its requirements were
based on the traffic loads, the spatial distribution and dynamics of
the daughterships, and onboard science data compression processing.

\begin{figure}[htb]
\includegraphics[scale=1.3]{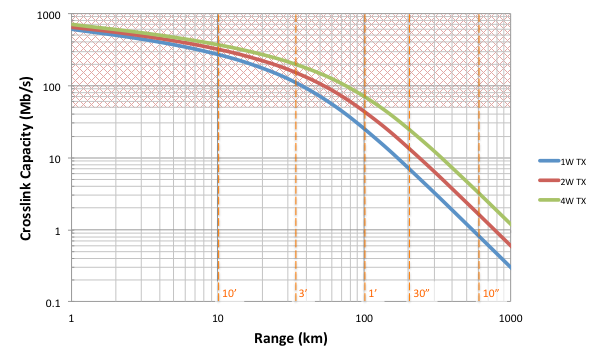}
\caption{Daughtership-to-Mothership crosslink communications data rate vs. range using UHF $\frac{1}{2}$-wave dipole antennas. Vertical lines indicate angular resolution for science array at different constellation spatial radii.}
\label{CoomDataRateVSrange:fig}
\end{figure}

In conjunction with topology selection, we conducted a study on different radio frequency bands (\hbox{UHF}, S~band, X~band, Ka~band) together with alternative antenna types (dipoles, parabolic, patch) and transmit power levels.
Figure~\ref{CoomDataRateVSrange:fig} shows an example for a daughtership-to-mothership communications link using a UHF half-wave dipole.
Assuming each daughtership transfers its science data directly to the centrally-located mothership, the radio range constrains the spatial extent of the constellation, which in turn impacts the resolution capability of the science array. 
Systems engineering considerations such as cost and feasibility of deploying different antenna types on cusbesat-class daughterships also had a significant influence on forming the resulting communications architecture.

Assuming a phased concept of operations, we found that the duration
required to crosslink the science data from the daughterships to the
mothership dominates the proportion of total mission operations (Figure~\ref{duration_for_science:fig}).
The baseline method to transfer daughtership science to the mothership consists of a series of scheduled daughtership-to-mothership crosslinks until all data is collected at the mothership; this is essentially a coarse Time Division Multiple Access (TDMA) approach. 
One possibility to reduce the total time required for this phase is to use Frequency Division Multiple Access (FDMA).
In this case, the mothership has the enhanced capability to receive multiple crosslink channels simultaneously. 
For cost and scalability, each daughtership is a smallsat having limited resource capabilities, which is maintained by the FDMA approach that imposes greater capability only on the mothership.
According to our operations concept, crosslink transfers will be scheduled to occur when the daughtership-mothership range is small in order to achieve maximum data rate capacity.
Because the daughterships are in motion, each daughtership radio must be able to re-tune its FDMA channel, since the set of daughterships in close range is dynamic (this is a very modest requirement).
Some technology development would be required to enable the mothership multi-receive capability, but it would be possible to implement a multi-user modem in an architecture developed by \citep{bskk14}.
It is also noted that sufficient spectral bandwidth will need to be allocated to accommodate the multiple crosslink frequency bands.
We also conducted an analysis to determine the potential benefit of using multi-hop communications among the daughterships to transport science data to the mothership.
That is, rather than a daughtership transferring its science data directly to the mothership, it may be advantageous to transfer it to another daughtership which will then relay it to the mothership (possibly via even more daughterships).
The potential benefit depends on the spatial distribution of the nodes, and arises from link performance being directly proportional to the square of the distance (in free space).

We take the approach of assuming each daughtership transmits at a fixed power but is able to adjust its data rate so as to meet a specified link margin.
For example, denoting $r_{ij}$ as the distance from daughtership~$i$
to daughtership~$j$ and~$r_{jM}$ as the distance from daughtership~$j$
to the mothership, then it will take less time to transfer the science
data from daughtership~$i$ via daughtership~$j$ whenever $r_{ij}^2 + r_{jM}^2 < r_{iM}^2$.
Based on this model, we determined improvements from using multi-hop routing by analyzing the spatial distribution of nodes at different snapshots of mission time, assuming a reasonable choice for the daughtership orbits.
Table~\ref{tab:multi-hop} presents results found.

\begin{figure}[thb]
\includegraphics[width=0.5\columnwidth]{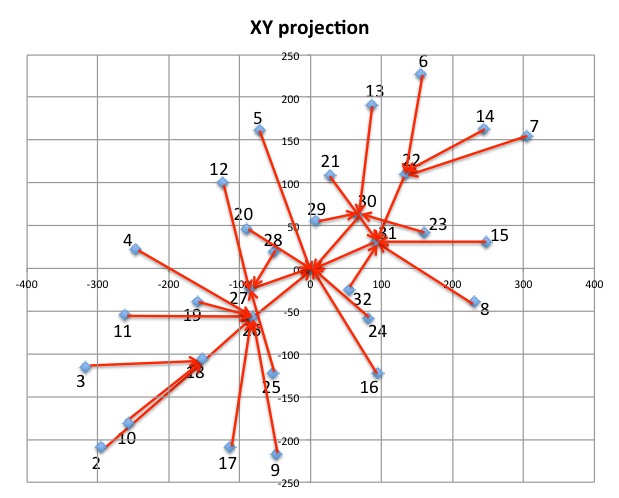}
\caption{Multihop paths to mothership, 40-day case.}
\label{Multihop_path_40days:fig} 
\end{figure}

\begin{table}[h]
\caption{Multi-hop Gains at Different Mission Times.}
\label{tab:multi-hop}
\centering
\begin{tabular}{|L|L|L|L|L|L|L|} \hline
{\bf Days into the mission}                         & {\bf max range from mom (km)}         & {\bf max internode range (km)} & {\bf \# of 2- or 3-hop paths $<$1-hop} & {\bf 2-hop time reduction} & {\bf \# of 3-hop paths $<$1-hop} & {\bf 2- or 3-hop time reduction}\\
\hline
15 days & 260.23 & 492.10 & 16 & 22.3\% & 2 & 22.9\% \\
\hline
30 days & 297.91 & 583.41 & 18 & 27.7\% & 4 & 31.9\% \\
\hline
40 days & 341.08 & 700.04 & 23 & 31.5\% & 6 & 36.7\% \\
\hline
54 days & 402.67 & 782.33 & 19 & 27.3\% & 4 & 30.0\% \\
\hline
mean & 325.47 & 639.47 & 19.0 & 27.2\% & 4.0 & 30.4\%\\
\hline
\end{tabular}
\end{table}
The maximum range to the mothership and between any nodes is included for a general sense of constellation span. 
A mean of 19 of the 31 daughtership-to-mothership paths are improved by mult-hop, 4 of which are 3-hop paths. 
No cases of 4-hop paths were found to provide benefit. 
The reduction in time to transfer all data is reduced by mean of 30.4.
An indication of the routing paths is depicted in Figure~\ref{Multihop_path_40days:fig}.
This corresponds to day 40 of the mission. 
A planar projection is shown, and thus link lengths may appear somewhat contorted. 
We see that certain intermediate nodes are selected to relay data on behalf of numerous ``outer'' constellation nodes.

Practical considerations of using multihop communications include the following:
\begin{enumerate}
\item Daughterships will need to communicate with one another at the ``return telemetry'' data rate.
Space radios typically operate using full duplex (i.e., different channels for transmission and reception) with different possible rates in each direction. 
(Typically ``command'' data rate is much less than the ``telemetry'' data rate.)
Even if the rates are symmetric, coordination of the inbound/outbound channels will be necessary.
An alternative is to use half-duplex (single-channel) communications, but full duplex is typically used to support ranging and Doppler measurements.
It is noted that full duplex may allow simultaneous reception (of data to be forwarded) and transmission, which would further speed the transport of data, however the analysis above did not assume this.
\item Selectable data rates will be constrained to some number of specific discrete values, and this quantization will yield some degradation of the results shown above.
\item There will be more energy consumed for daughtership relay receptions.
\item Certain ``inner'' daughterships will bear the burden of relaying data from numerous ``outer'' daughterships, in addition to its own science data, so that the work is unevenly distributed.
\item One might consider orbit selection that also accounts for improved multihop performance; currently we have focused primarily on science quality.
\end{enumerate}

\subsection{Data rate discussion}\label{datarate:subsec}
We considered the following options for the data downlink:
\begin{itemize}
\item raw waveform downlink
\item using polyphase filters
\item Fast Fourier Transform, FFT, on board
\item 1, 3 and 12 bit signal digitization, $N_{dig}$.
\end{itemize}
We also assume 2:1 lossless compression ratio in addition to any data reduction technique used.
For this discussion, we only consider the transmission at the carrier frequency.
The data rate for each case is described below.

\subsubsection{Raw data downlink}\label{rdd:sec}
The raw data downlink assumes that the recorded waveform is
transferred to the mothership continuously. In the calculations below,
we assume the highest frequency of the system to be 30~MHz, requiring
$f_N=60$~MSps digitization speed. For a 1-bit digitization, the
resulting data rate is 60~Mbps per daughtership for two
polarizations. We assume a factor of 2 lossless compression of the
signal. The data transfer rate grows linearly with the digitization
depth, and, for 12-bit digitization, equals 720~Mbps. An array of~32
daughterships would require about~24~Gbps downlink.

\subsubsection{Polyphase Filters}\label{pf:subsec}
The basic idea of polyphase filters is the splitting of the  channel into multiple narrow band channels with subsequent digital heterodyning of each channel.
For instance, a 16~kHz wide channel would  require only 32~kSps digitization speed.
The narrow-band channels are later combined reconstructing the original signal with minimal loss. 
More details on the subject can be found in~\cite{porat}.


The use of polyphase filters allows to reduce the data rate by a factor of~60; in the case of 12-bit digitization, the data would be
about~12~Mbps per daughtership or about~384~Mbps for the whole 32-daughtership constellation.

\subsubsection{FFT Onboard the Spacecraft}\label{fft:subsec}
The idea is to do fast Fourier transform of the recorded waveforms,
FFT, on board the individual daughtership and downlink the FFTs
instead of the raw waveforms in order to perform the cross-correlation
either on board the mothership or on the ground.  This approach would
reduce substantially the data rate required.


If the number of required frequency samples in the \hbox{FFT}, $N_{freq}$, is taken to be 64 and if $N_{time}$ is the number of samples in the time domain, then the FFT would have to have $\frac{1}{2}N_{time}$ samples. In order to obtain $N_{freq}$, a minimum $N_{time}=128$ samples would be required; for the purpose of averaging, the next higher number should be used, which would be 256 in this case.
 
The minimum integration time required to obtain $N_{\mathrm{freq}}$ is $t_{\mathrm{int}}=\frac{N_{\mathrm{time}}}{f_{N}}=\frac{128}{60\times 10^6}\approx2\times10^{-6}$~s.
We cannot have integration time less than this value if we aim to obtain at least 64 samples in frequency domain.  Assuming continuous observation, the downlink data rate is $R=N_{\mathrm{freq}}\times\frac{N_{\mathrm{bits}}}{t_{\mathrm{int}}}\approx350$~Mbps, or~11~Gbps for an array of~32 daughterships, assuming 2:1 lossless compressing offseting two polarization signals per daughtership.
The data rate calculated above represents a lower limit of the integration time. 
The usual internal integration time of the receivers is about~1~s, meaning the real data rate will be about 6 orders of magnitude lower and the minimal integration time condition fulfilled without any problem.

\subsection{Ranging system requirements}
In order to provide initial condition for science data processing and radiometrics for orbit determination, position knowledge and time derivatives must be obtained for each daughtership. We discuss the effect of the time derivatives in Section~\ref{sec:image} and only consider the baseline vectors here.
In order to reduce cost and improve
operational scalability, position knowledge is derived from a
two-tiered process. On the top tier is the mothership, for which the
absolute position is determined by standard DSN procedures.  At the second tier, the relative position of each daughtership to the mothership is derived from a combination of gravity model and crosslink range measurements performed within the constellation.

The desired upper bound on position knowledge uncertainty is~30~m,
roughly equal to the wavelength at~10~MHz. Assuming that each
daughtership's relative position can be computed through
multi-lateration of range estimates with one-third of the
constellation nodes, i.e., 10 other daughterships, simulation shows
that a range estimation error no greater than~1~m is desired. In terms
of navigation, the minimum safe separation distance is~3~km, which is
much larger than the~30~m science requirement that will form the basis
of our ranging system design and analysis. Due to daughtership motion
and expansion of the constellation orbits, it is estimated that at a
minimum inter-spacecraft range should be measured every 10 to~15
minutes. Although there are 496 unique bi-directional
inter-daughtership crosslinks at any given time, in order to derive
sufficient position knowledge, we expect that each daughtership will
need to range with approximately 10 other daughterships, reducing the
number of measurements below the worst case of~320---the exact number
required depends on the geometry and how many redundant measurements
can be eliminated, as the range estimate from daughtership~$i$ to~$j$
is symmetric with daughtership~$j$ to~$i$. Assuming an integration time of~1~s for each measurement, it takes 320~s to complete one cycle of ranging measurement. In reality, it will take much less than that time since ranging measurements with sufficient spatial separation can be conducted simultaneously.

\subsubsection{Strawman Design of Crosslink Ranging System}

In designing a crosslink ranging system, we attempt to maximize dual-use of the relay communications system's transceiver and antenna to save cost, weight, and most importantly space on the daughtership. Currently we assume that the mothership will utilize the same relay system as the daughtership although it is definitely possible for the mothership to carry a compatible yet far more capable system in flight. For our strawman design, however, we assume that the relay system is identical on both the mothership and the daughterships. We assume that the relay system may operate around 500~MHz (UHF) or 8.4~GHz (X-band). For UHF operation, we consider a single dipole antenna extending along the $Z$ axis at half-wavelength or $\frac{5}{4}$ wavelength, the latter having higher directivity but narrower beamwidth. For $X$-band operation, four pairs of transmit and receive patch antennas mounted on the four sides of the daughtership, facing the $+X$, $-X$, $+Y$ and $-Y$ axis directions, provide higher gain than dipole and wide coverage. We consider these two antenna initially to minimize pointing requirements. As we will describe later, a horn antenna option was also considered with increasing operational complexity due to pointing requirements.

In terms of ranging technique, we consider PN ranging using either coherent (turn-around) mode or regenerative mode. Sequential tone ranging require synthesizing different carrier and good knowledge of starting time, both are difficult to implement within a constellation of smallsat. Therefore, only PN ranging is considered in our study. Coherent mode has the advantage of eliminating frequency mismatch between the uplink and the downlink at the cost of requiring a transponder, while regenerative (non-coherent) mode achieves better economy by using a transceiver and provides some degree of performance advantage under high noise environment. As the relay system design trade is still open, both coherent and non-coherent modes are considered in our analysis. To analyze ranging system performance, we used two PN codes \cite{DSN}, one designed for coherent and the other for non-coherent ranging, and assume a nominal ranging clock of 1~MHz. For turn-around ranging, the noise equivalent bandwidth is 1.5~MHz. Since both codes have an ambiguity resolution of 75,660~km \cite{DSN}, far greater than the size of the constellation orbit with maximum baseline of 600~km, they are suitable for our purpose.

We assume that the ranging signal will utilize squarewave, and the transponder?s automatic gain control (AGC) delivers constant root-mean-square (RMS) voltage. We do not consider the more complex scenario of simultaneously ranging and carrying command and telemetry on the crosslinks, therefore all transmit power is allocated to the ranging signal component. Although under good SNR environment there is no reason not to exchange command and telemetry while ranging and there is good potential for reducing latency, it is the beyond the scope of this initial analysis.

The root-mean-square error of the ranging measurement is given by \cite{DSN}
\begin{equation}
\sigma=\frac{c}{\sqrt{f_{\mathrm{RC}}A_CR_1\sqrt{256T\frac{P_R}{N_0}|_{D/L}}}},
\label{EQ_Ranging1}
\end{equation}
where $f_{\mathrm{RC}}$ is the frequency of the range clock, $A_C$ is
the fractional loss
of correlation due to frequency mismatch (0 for coherent mode with
assuming Doppler rate aiding~\cite{Berner}; $|\sin c(2 f_{\mathrm{RC}} T)|$ for
regenerative mode, where $f_{\mathrm{RC}}$ depends on oscillator accuracy and
stability, with no Doppler rate aiding assumed), $R_1$ is cross-correlation
factors for the PN code, $T$ is the integration time (in seconds), and
$\frac{P_R}{N_0}|_{D/L}$ is the ranging signal signal-to-noise ratio on the
downlink. For turn-around ranging, the uplink noise has a much higher
contribution to the downlink signal-to-noise ratio than regenerative ranging. A detailed
description of the relationship between the uplink and downlink SNRs
and their respective contribution to the ranging error can be found in
\cite{DSN}.  Table~\ref{tab:croslinacc} summarizes the required $\frac{P_R}{N_0}$ for achieving 1~m accuracy, as derived
from equation~\ref{EQ_Ranging1}.

\begin{table}[h]
\caption{Crosslink Ranging Accuracy and SNR Requirements.\protect\footnote{Patch antenna parameters: dielectric constant of substrate = 2.2, height of substrate is 0.16~cm, width is 1.41~cm, length is 1.1~cm, resonant frequency is 8.4~GHz}}
\label{tab:crosslinacc}
\centering
\begin{tabular}{|c|c|c|c|c|c|} \hline
Ranging Technique                          & Received Pwr         & Required $\frac{P_R}{N_0}$ & Ranging Accuracy & Integr. & Antenna Type \\
                                                        & (uplink, W)                  & (downlink)                              & (m, RMSE)            & Time (s)                      &        \\ 
                                                        \hline
Turn-around (UHF)                           & $1.09\times10^{-14}$ & $1.05\times10^{6}$                & \begin{tabular}{@{}c@{}}0.05 \\ 0.016\\0.006\end{tabular} & \begin{tabular}{@{}c@{}}1\\ 10\\100\end{tabular}&\begin{tabular}{@{}c@{}}$\frac{1}{2}$wave dipole, 2.1 dBi \\ 1.25 wave dipole, 5.1 dBi\end{tabular} \\
\hline
Regenerate, non-coherent (UHF)     & $1.09\times10^{-14}$ & $2.13\times10^{6}$                &  \begin{tabular}{@{}c@{}}0.25 \\ 0.23\end{tabular} & \begin{tabular}{@{}c@{}}1 \\ 2\end{tabular}& \begin{tabular}{@{}c@{}}$\frac{1}{2}$wave dipole, 2.1 dBi \\ 1.25 wave dipole, 5.1 dBi\end{tabular} \\
\hline
Turn-around (X-band)                       & $1.30\times10^{-15}$ & $4.82\times10^{2}$                & 1                           & 6                         & \begin{tabular}{@{}c@{}}conical hornm 11~cm \\ aperture, 17.07 dBi\end{tabular}\\
\hline
Regenerate, non-coherent (X-band) & $1.36\times10^{-16}$ & $2.90\times10^{4}$                & 1.45                     & 1                         &  patch (1.4 $\times$ 1.1~cm), 7.2 dBi \\
\hline
\end{tabular}
\end{table}
\footnotetext{Patch antenna parameters: dielectric constant of substrate = 2.2, height of substrate is 0.16~cm, width is 1.41~cm, length is 1.1~cm, resonant frequency is 8.4~GHz}
Table \ref{tab:crosslinacc} shows that to achieve 1~m accuracy, 1~sec integration time will be sufficient except in the case of the X-band turn-around ranging, where a horn antenna and more accurate pointing will be required. The next step in our analysis is to apply the required $\frac{P_R}{N_0}$ threshold when simulating the orbital geometry and antenna pointing in order to collect statistics on ranging opportunities in the natural course of a two-month long observation campaign. 

\subsubsection{Constellation Ranging System Performance Analysis}

We utilized binary SPK kernels
developed during the orbital design and generated daughtership orientation kernels that help model pointing and compute antenna and received power levels. Due to the constant motion and the need for orbit maintenance, we assume that ranging could occur during both science collection and data relay phases. For the initial two-month long scenario, we assume for sake of simplicity that the primary orientation for all daughterships is to have the $+Z$ axis pointed in the direction normal to the orbital plane of the mothership, and that for the secondary orientation the $+X$ axis is pointed toward the Sun to maximize power generation. In reality, we expect the constellation network pointing to be different depending on whether it is optimized for science or relay. However, we make this simplifying assumption in order to quantify, at a coarse grain level, the temporal characteristics of the ranging links.

\begin{figure}[thb]
\includegraphics[width=0.9\columnwidth]{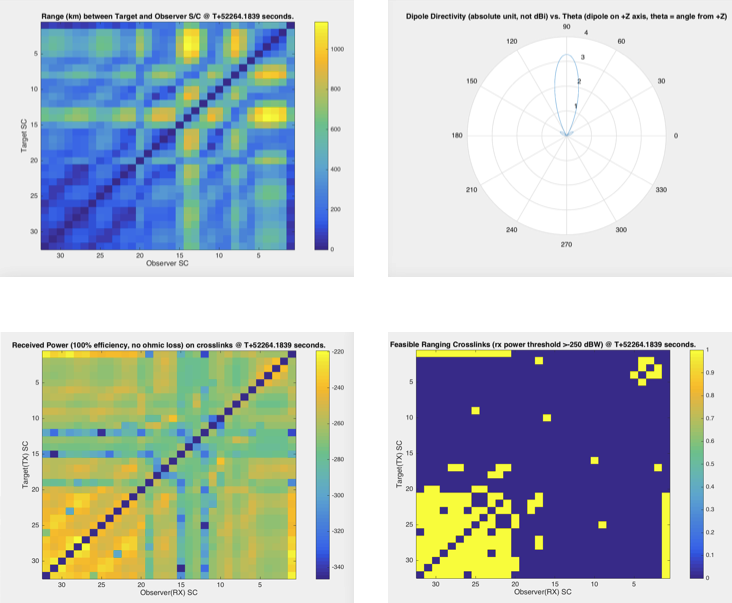}
\caption{An example of constellation crosslink geometry simulation. Upper left: Range between the target and the observer. Upper right: dipole directivity used for the simulation. Lower left: power received assuming 100\% efficiency. Lower right: feasible ranging cross-links assuming 250 dBW power threshold.}
\label{MATLAB_SPICE_model_screen_shot:fig} 
\end{figure}

This model incorporating the daughtership trajectory, orientation, antenna radiation pattern, and ranging link analysis is used to simulation the geometry and ranging performance of the constellation network. Figure~\ref{MATLAB_SPICE_model_screen_shot:fig} is a screen shot of some of the metrics generated by the simulation such as the antenna pattern, the crosslink range and range rate, the received signal power, and the feasible ranging links meeting certain SNR threshold. In our simulation study, we assume a transmit power of 4 watts, Allen deviation of the oscillator to be $1\times 10^{-9}$~Hz/s, and noise temperature of 290K + 50K of cosmic background. The simulation focused on three configurations: (a) UHF band turn-around ranging with half-wavelength dipole, (b) UHF band turn-around ranging with 1.25 wavelength dipole, and (c) X-band regenerative ranging with 4 pairs of patch antenna. For dipoles we use computed antenna patterns in our model; for patch antenna, we use a ``cookie-cutter'' model in which half the maximum antenna gain is assumed when a target is within the 3dB beamwidth and all targets outside the beamwidth are assume unreachable. Those configurations not covered by our simulations, namely, UHF with regenerative ranging and X-band turn-around ranging, are still open options to be considered in future studies. 

\begin{figure}[thb]
\includegraphics[width=0.7\columnwidth]{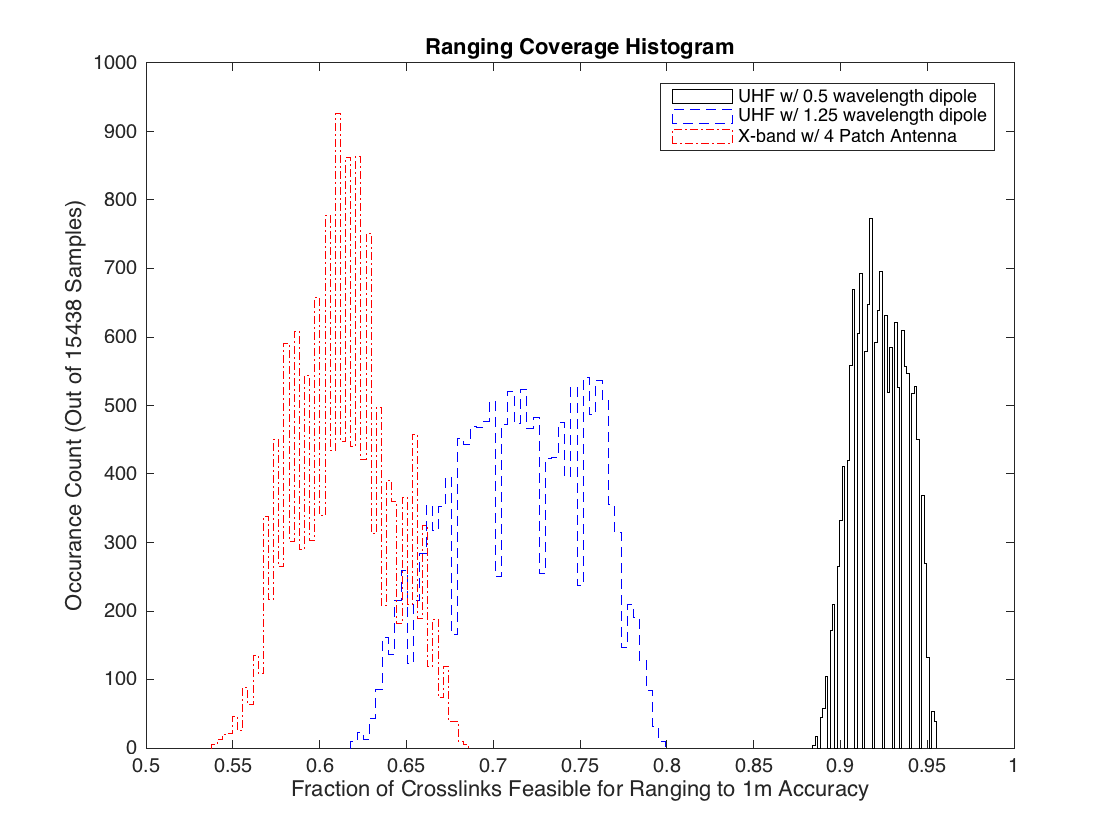}
\caption{Histogram of the Fraction of All Crosslinks Achieving 1m Ranging Accuracy Per Time Instance}
\label{Ranging_Coverage_Histogram:fig} 
\end{figure}

Figure \ref{Ranging_Coverage_Histogram:fig} shows the histogram of the fraction of all crosslinks that is able to achieve 1m accuracy. The simulation covers 15,438 5-minute time steps, approximately 54 days of operation. At each time step, the ranging signal SNR is computed for all possible crosslinks, taking into about antenna pointing, radiation pattern, and space loss, and the percentage of crosslinks that can achieve 1m accuracy is recorded in the histogram. The result shows that our baseline constellation geometry achieves excellent ranging coverage for all three configurations considered. The half-wavelength dipole at UHF has the best coverage such that at all times more than 90\% of all crosslinks can deliver 1m accuracy ranging. Upgrading to the 1.25 wavelength dipole provides 3dB increase on antenna directivity but reduces the number of feasible ranging links to an average of 70\%. For X-band using 4 pairs patches, on average more than 60\% of the crosslinks are feasible for conducting ranging measurements.

\begin{figure}[thb]
\includegraphics[width=0.7\columnwidth]{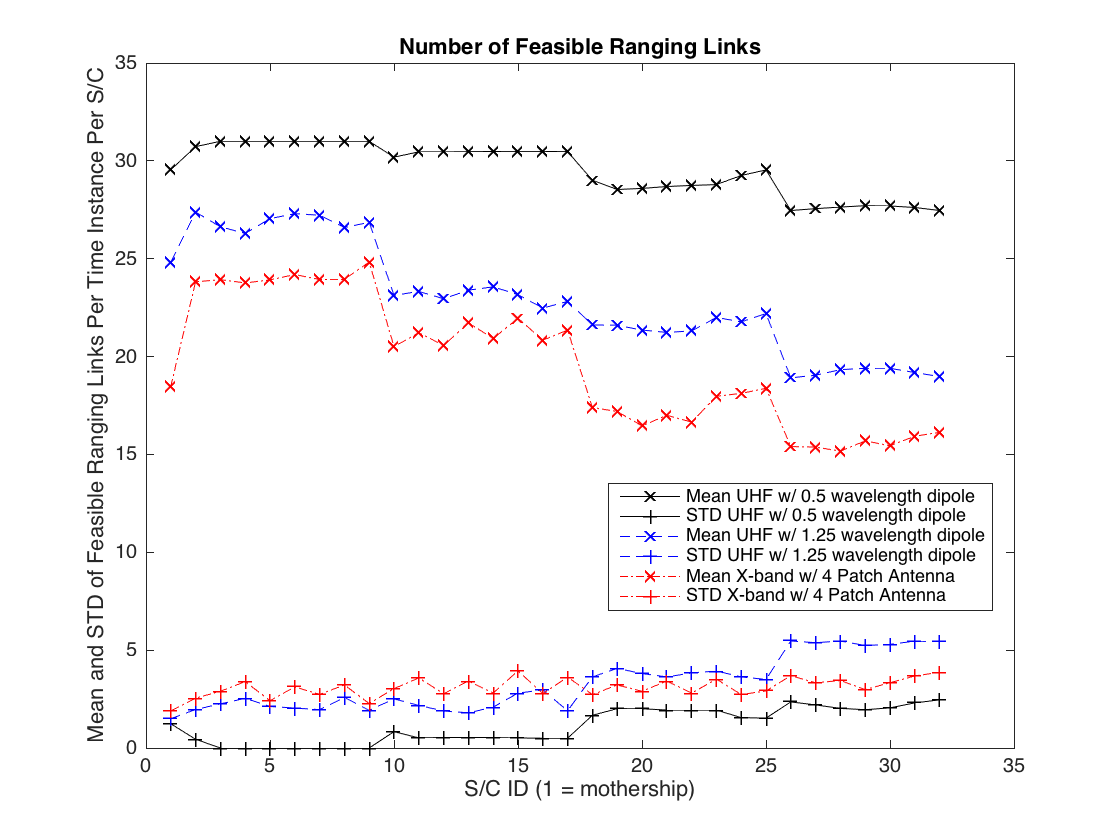}
\caption{Number of Feasible Ranging Links per Daughtership per Time Instance.}
\label{Number_of_Feasible_Ranging_Links:fig} 
\end{figure}

In order to better understand how many ranging measurement can be done
per daughtership with a short time window, we need to further analyze
the distribution of the feasible ranging links among the 32
daughterships. Figure~\ref{Number_of_Feasible_Ranging_Links:fig} shows
the mean and standard deviation of the number of feasible ranging
crosslinks for each daughtership sampled over 15,438 time instances.
The same general trend persists, whereby the half-wavelength dipole
achieves the highest average number of feasible ranging links and
lowest standard deviation.  As might be expected, the ranging
opportunity would be better for those daughterships in the inner rings
(daughterships~1 to~9) than those on the outer rings. For the worst
case, the outer ring with X-band crosslinks, there would be an average
of~15 feasible ranging links per daughtership, with a standard
deviation of~4, which would be acceptable.   Even for daughterships in
the outer ring, it is not difficult to achieve 30~m position
knowledge. Given the short integration time of only 1~s, we do not
anticipate that there would be difficulty in providing timely ranging estimate updates to support both science processing and orbit maintenance.

One of the outstanding issues that has not been adequately addressed
in our preliminary study is interference avoidance while
ranging. Further analysis is required to determine the required
spatial and angular separation when several crosslinks conduct ranging
operations simultaneously. In that context, the benefit of
implementing FDMA or 802.11-like CSMA medium access control (MAC)
protocols for coordinating the crosslink communications and ranging
can also be evaluated. As the mission concept is further refined, it
should become possible to incorporate a more detailed daughtership
orientation profile into the ranging simulation. Along that same line
of investigation, a more detailed trade study between using high gain
antenna and its effect on pointing and scheduling of science, relay,
and ranging operations.  Finally, there is the concept of simultaneous command/telemetry transmission and ranging; the potential saving in latency and power should be quantified in terms of impact on ranging accuracy. Recent advanced in telemetry-based two-way ranging \cite{TLMRNG} research shows great potential in reducing range jitter by taking advantage of high rate telemetry signal structure. These advancements in ranging research should be also incorporated into future studies.

\section{Strawman Spacecraft}\label{sec:spacecraft}

In order to form an interferometer, all of the signals from the
individual antennas must be brought together for \emph{correlation}.
In ground-based telescopes, the correlation occurs in a dedicated processing centers.
For the RELIC design, we
have retained this architecture, which necessitates a
{mothership} to which {daughterships} send the collected
science signals.  Because of their different functions, the
daughterships and the mothership cannot be identical, and we describe
each in turn.  Our focus is on the daughterships, as key elements of
the telescope---the reception of radio waves and the position
determination---are associated with the daughterships, while the
mothership design is more straightforward.

\subsection{Daughtership Design}\label{sec:spacecraft.daughter}

As described above, the required effective aperture exceeds 100~km
leading us to adopt an interferometric design for the RELIC mission.
This design choice allows us to leverage a significant knowledge base
\cite{tms07} and experience from ground-based interferometric
telescopes, including many that operate at frequencies not too
dissimilar than those of RELIC
\cite{bmrsz78,s90,a95,vla74,lofar,lwa1,tek+12,mwa}.

A natural approach to an interferometric telescope is for it to
consider of a number of identical daughtership spacecraft, with each daughtership
carrying a single science payload consisting of an antenna and
receiver.  This approach is what has been adopted in previous mission
concepts \cite[e.g.,][]{fhr67,wjs+88,alfa,op05,bljsfa13}, and it is a
straightforward extension of how ground-based interferometers are
designed.  The science payload itself is relatively simple, the
daughtership would be exposed to essentially identical environmental
conditions allowing them to be identical, and the design can leverage
economies of scale in constructing multiple identical spacecraft.

\begin{figure}[!b]
    \begin{center}
       \includegraphics[width=0.75\textwidth]{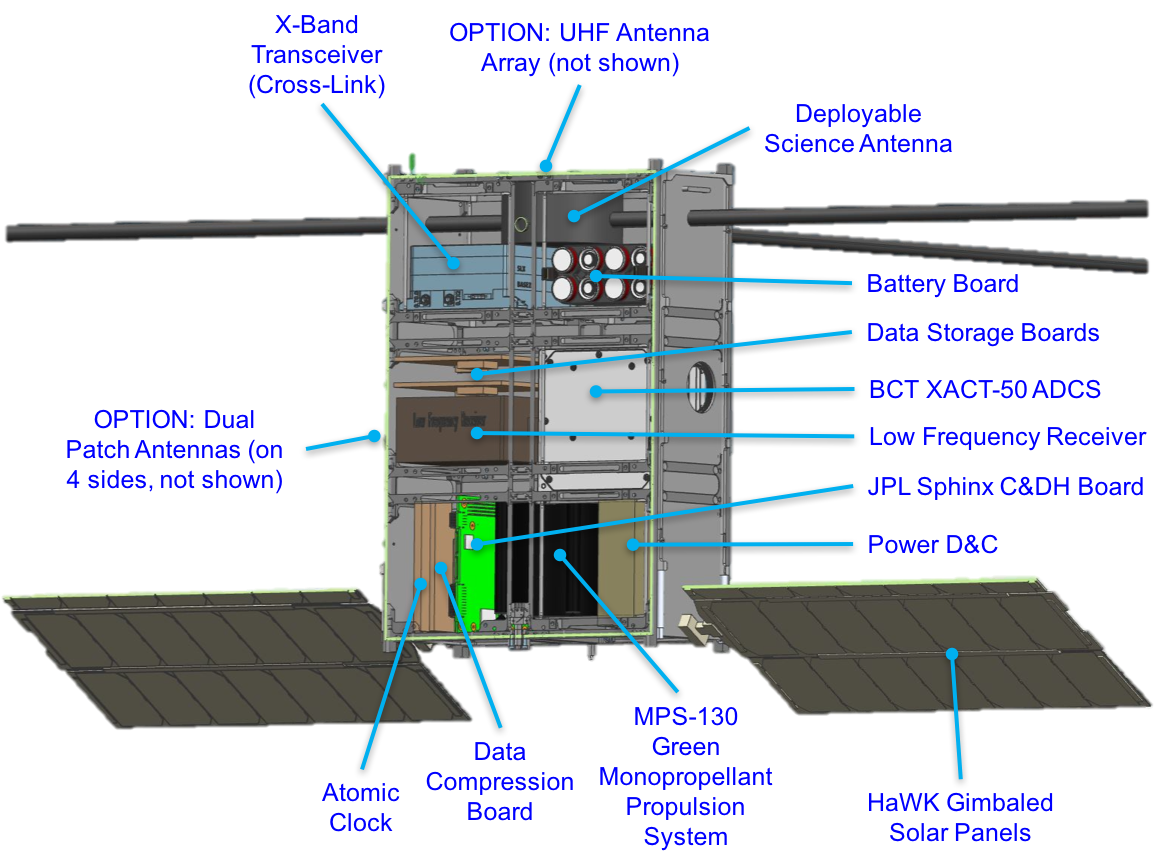}
       \caption{Physical layout of the RELIC daughtership spacecraft. Cabling is not shown.}
    \end{center}
\label{fig:daughtershipDesign} 
\end{figure}

The daughterships are designed to conform to a 6U form factor
($10\,\mathrm{cm} \times 20\,\mathrm{cm} \times 30\,\mathrm{cm}$) as
the science payload itself is not large and this form factor allows
significant leveraging of compact and standard components, thereby
likely minimizing overall mission cost.
Table~\ref{tab:daughtershipProperties} summarizes key high-level parameters
of the daughtership design, and Figure~\ref{fig:daughtershipDesign}
presents the physical layout of a daughtership.  The remainder of this
section presents details of specific sub-systems within a daughtership.

The daughterships would be deployed from a container (``pod''), which
secures the daughterships during launch and  transportation to the
final orbit.  The ``pod'' also would provide initial momentum for reaching the desired orbital configuration.

\begin{table}[h]
\caption{Daughtership Parameters}\label{tab:daughtershipProperties}
\centering
\begin{tabular}{|l|l|} \hline
\textbf{Property} & \textbf{Value (Incl. Contingency)} \\ \hline
Mass & $<$ 14~kg \\ \hline
Peak Power Consumption & 52~W \\ \hline
On-Board Data Storage Capacity & 32~GB \\ \hline
Form Factor & 6U ($10\,\mathrm{cm} \times 20\,\mathrm{cm} \times 30\,\mathrm{cm}$) \\ \hline
Pointing Accuracy & $< 0.03\deg$ \\ \hline
Slew Rate & $10 \deg/s$ \\ \hline
Maximum Thrust & $>$ 1~N \\ \hline
Available $\Delta V$ & $>$ 100~m~s${}^{-1}$ \\ \hline
Maximum Power Generation (beginning of life) & 62.5~W \\ \hline
Battery Capacity & 77~W~hr \\ \hline
Expected Unit Cost (incl.\ payload) & \$5M (TBC) \\ \hline
\end{tabular}
\end{table}

\subsubsection{Science Payload Subsystem}\label{sec:spacecraft.daughter.science}

The science payload consists of a dual-polarized antenna, receiver,
and associated signal processing unit capable of receiving radio
frequencies between~0.1~MHz and~30~MHz.  Following standard practice, we
adopt an electrically short dipole, as the system temperature is
determined entirely by the synchrotron emission of the Galaxy
\cite{c79} and dipoles approaching resonance are large
enough to be potentially mechanically unstable \cite{akngw75}.

For this design, a motorized, deployable dual dipole antenna has been
chosen.  Dipole antennas have been developed that, when stowed, can
fit within a volume of~\hbox{0.5U} and, upon deployment, have an effective length for each of the four poles of~1.2~m.
More details about the antenna design are provided in Appendix A. 


The receiver and signal processing unit amplify, condition, and sample
the incident radio waves.  Designs for such units are well developed
\cite[e.g.,][]{wind,cassini,stereo,juno}.  For ground-based systems,
analog signals are digitized with anywhere from~1~bit to~24~bits per
sample.  For a space-based array such as \hbox{RELIC}, data
transmission is a significant factor and high bit depths can lead to
infeasible data rates.  We have baselined 12~bits per sample to enable
sufficient bit depth to handle the range of powers expected over the
relevant frequency range.  We further employ a polyphase filter bank
to produce a set of frequency sub-bands \cite{kgb+15}
The resulting effective data rate for science data collection is 12~Mb~s${}^{-1}$ per daughtership.


The overall power usage of the science payload subsystem is expected
to be less than 10~W in operation and less than 1~W when the device is in standby.

\subsubsection{Communications Subsystem}\label{sec:spacecraft.daughter.comm}

The communication subsystem is required not only to transmit data from
the daughterships to the mothership but also to conduct ranging
between neighboring daughterships in order to determine relative
positions for the interferometry.  A key trade for the communications
subsystem is to minimze the power required while maximizing the data rate.


Two, nearly equivalent options exist for the communications subsystem.
Option~1 is an ultra-high frequency (UHF) system operating at~450~MHz,
using a deployable, canted turnstile antenna with a gain of~1~dBi
at~450~MHz and with an operational frequency range
of~400~MHz--480~MHz.  Option~2 is an X-band system operating
at~8.5~GHz.  For this design, a set of four pairs of high-gain patch
antennas tuned to~7.4~GHz (receive) and~8.5~GHz (transmit) and a gain
of~12~dBi, each mounted to a 10~cm $\times$ 10~cm surface on four
sides of the daughtership was selected.

Both options use approximately 5~W of power to transmit the required
data rates.  The X-band option is based on marginally less mature
technology for small spacecraft, but it brings the advantage over UHF
of being more compact and carrying less risk of possible interference
with the science payload.  Conversely, the UHF option is likely to be
less costly and less complex (particularly since only one antenna
array is required).  

In order to allow for sufficient link margin, design parameters of a
16~Mbps at~100~km distance from the mothership were used.

One potential complication is that requirements on allowable
frequencies for inter-satellite links in lunar orbits may not be
satisfied using either X- or UHF-band solutions.  A waiver may be required to implement either of these options.


\subsubsection{Command and Data Handling Subsystem}\label{sec:spacecraft.daughter.c&dh}

Each daughtership has a command and data handling (C\&DH) subsystem, consisting of a data processing unit and data storage. The requirements on the data processing unit are modest, in that the flight computer is primarily to be used for navigation and ranging. Data storage is primarily to be used for buffering data to be downlinked.

As an initial design for \hbox{RELIC}, the radiation-tolerant flight
computer Sphinx, developed at \hbox{JPL}, was selected.  The Sphinx
fits within the allowed 6U form factor and  features a reprogrammable field programmable
gate array (FPGA) and a Leon3 dual core processer clocked at 100~MHz.
The Sphinx flight computer board has 8~GB of radiation tolerant (192~krad total dose) non-volatile memory, and features a number of interfaces, including a high-speed dual SpaceWire adapter capable of supporting up to~200~Mb~s${}^{-1}$. The latter is required during the cross-link phase to forward communication packets to the communication subsystem at desired speeds.

Data collected by the science payload on the daughterships could not be cross-linked to the mothership simultaneously,
particularly due to the large number of daughterships. Therefore, a
buffer is required.  The amount of memory on the Sphinx flight
computer is not sufficient to store the data from a typical RELIC
observation (4~hr generating 22~GB).  Allowing for additional
spacecraft housekeeping data, error correction checksums, and
contingency, a total of~32~GB is needed.  This data storage volume is
met by using four 8~GB memory modules (e.g., radiation tolerant FLASH
NAND). An appropriate board housing these memory chips, and providing
an appropriate high-speed data interface and controller do not yet
exist and would need to be designed.

\subsubsection{Attitude Determination and Control Subsystem}\label{sec:spacecraft.daughter.adcs}

In order to orient the daughterships to conduct the science
observations and for power generation for the solar panels, the
daughterships are equipped with an attitude determination and control
subsystem (ADCS). Requirements with respect to pointing accuracy for
the science observations are modest, estimated to be 0.1\deg, given
the large power patterns provided by the electrically short dipoles.

Required for the RELIC daughterships are at least three reaction
wheels for 3-axis control, an equal number of torque rods to decrease
$\Delta v$ requirements for desaturation, and star trackers for
positional determination.  A plethora of off-the-shelf components for
ADCS exist that would meet the requirements.


\subsubsection{Propulsion Subsystem}\label{sec:spacecraft.daughter.propel}

The mission concept of operations specifies a number of maneuvers that
require a propulsion system. These maneuvers are initial configuration
of the constellation, from drop-off from the transfer vehicle to the
initial formation; regular station keeping; and reconfiguration
maneuvers that increase the number of baselines.  A 3-axis propulsive
control system is required for desaturation of the reaction wheels due
to the long mission duration.  Table~\ref{tab:deltaVRequirements}
summarizes these different propulsion requirements, as currently
understood.

\begin{table}[h]
\centering
\caption{Maneuvers and Corresponding Required $\Delta v$}\label{tab:deltaVRequirements}
\begin{tabular}{|l|c|} \hline
\textbf{Maneuver} & \textbf{Required $\Delta v$} \\
	        & \textbf{(m/s)} \\ \hline
Initial Configuration & 20 \\ \hline
Reconfiguration (20 [TBR] times) & 1 (per reconfiguration) \\ \hline
Station Keeping & $<$ 1 (per day) (TBR) \\ \hline
Desaturation & TBD \\ \hline
\end{tabular}
\end{table}

There are a number of types of propulsion systems available for small
spacecraft, which can be classified broadly into chemical (cold vs.\
warm gas, mono- vs.\ bi-propellant.), electrical (ion, Hall-effect),
and propellantless (solar sail.) drives. Propellantless and electrical
drives have very low maturity and produce little thrust.  Cold and
warm gas drives have low energy density fuels, requiring large
quantities to satisfy the $\Delta v$ requirements.  A number of mono-
and bi-propellant drives exist, capable of providing
upto~100~m~s${}^{-1}$ of total $\Delta v$, 3-axis control, and not
requiring toxic fuel (thereby avoiding the need for a waiver).  



\subsubsection{Power Subsystem}\label{sec:spacecraft.daughter.power}

\begin{figure}[h]
    \begin{center}
              \subfigure[Power consumption over time for an individual daughtership.] {\includegraphics[width=.49\textwidth]{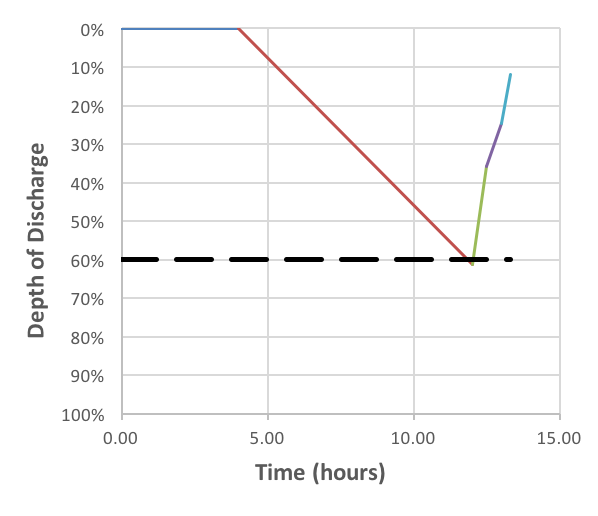}}
              \subfigure[Power profile for core mission activities.] {\includegraphics[width=.49\textwidth]{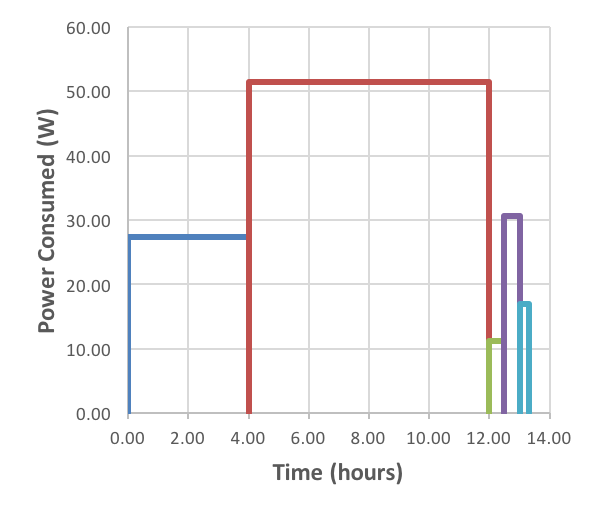}}
        \caption{Representative excerpts of the expected power profile and power consumption over time, including 30\% contingency. The sole power negative mode (red) is related to cross-linking data.}
    \end{center}
\label{fig:daughter.power} 
\end{figure}

Due to the desired high data rates and required data volume, power requirements are higher than is typical for small spacecraft.
Table~\ref{tab:daughter.powerModes} summarizes the mission profile,
which is illustrated in Figure~\ref{fig:daughter.power}, that was used
in order to design the electrical power system (EPS), including determining the size of the solar panels and batteries.
Most power modes assume a number of instruments being on (including
heaters), and each power mode assumes a 30\% contingency.

\begin{table}[h]
\centering
\caption{Primary Mission Power Modes and Durations}\label{tab:daughter.powerModes}
\begin{tabular}{|l|c|c|c|c|c|} \hline
\multirow{2}{*}{\textbf{Mode}} & \multicolumn{3}{|c}{ } & \multicolumn{2}{|c|}{Propulsive Maneuver} \\ \cline{2-6}
 & Cruise / Idle & Instrument Ops & Cross-Link & Pre-Heating & Firing  \\ \hline
\textbf{Power Usage (W)} & 11     & 28 & 52 & 31 & 17 \\ \hline
\textbf{Duration (hr)}  & \ldots &  4 &  8 & 0.5 & 0.3 \\ \hline
\end{tabular}
\end{table}


Solar panels consisting of~12 coplanar strings, each with~7 cells
generating approximately 0.8~W per cell at the beginning of life
(BOL), would generate 62.5~W at BOL and satisfy the power
requirements.  The strings would be attached to two gimbaled,
deployable solar panels similar to the HaWK array used by the JPL
mission \hbox{MarCO}. The sizing was based on the power modes, as well
as a (conservative) assumption of an eclipse fraction of~40\%
throughout most modes, with 25\% being assumed for data transmission
(pointing towards Sun whenever possible).  A worst case incident angle
of~10\deg\ was also assumed.

Data transmission can be particularly power intensive, and the
daughterships may undergo eclipses.  A lithium-ion battery solution,
consisting of eight batteries producing a total of~77~W~hr would be
sufficient for the assumed mission profile.  A maximum of~60\% depth
of discharge was assumed in determining the number of batteries
required.

\subsubsection{Thermal Management}\label{sec:spacecraft.daughter.thermal}

Primarily of concern in terms of thermal management would be the heat
produced by the (X- or UHF-band) transmitter and ensuring that the
temperature of both the propulsion unit and battery array would remain
within operating limits. Given the volumetric constraints of a small
spacecraft, primarily passive heat management would be
employed, with radiators being placed on the external structure.  Both
the structure and secondary heatpipes would ensure the flow of energy
to the radiators. Active heaters would be used for the battery array
and propulsion system in order to maintain their temperatures within
operating limits, but, in general, components would placed in such a
fashion that use of active thermal management units would be
minimized. For instance, the battery array could be placed close to
the transceiver, such that a portion of the heat generated by the
transmitter could flow to the battery array over the structure.

\subsubsection{Radiation and Shielding}\label{sec:spacecraft.daughter.radiation}

For the proposed mission, the radiation requirements would be fairly
modest, at least during nominal mission operations.  Previous measurements have shown an average dose of~10~krad~yr${}^{-1}$ to be expected in the selected orbit.
Most currently available small spacecraft hardware is qualified to
50~krad, and most of the assumed RELIC electronics are capable of
enduring a dose of up to~200~krad without additional shielding.  For a
mission duration of approximately 1~year, as well as the shielding
provided by the structure, radiation is unlikely to be an
issue. During transit to the final orbit, when the van~Allen belts are
traversed, additional shielding is provided by the transport craft,
the deployment ``pod,'', and the undeployed solar panels.

\subsection{Mothership Design}\label{sec:spacecraft.mother}

The mothership serves to accumulate the data streams from the
individual daughterships and transmit them back to Earth.  Primary
drivers for the design of the mothership were the data storage and the
communication subsystem.  Our objective was to design a mothership as
a small spacecraft with a mass of no more than 100~kg.  Most of the
subsystems have similar requirements as for the ones for the
daughterships and would be similar.  Key differences between the
mothership and daughterships would be the telecommunications and data storage.

The telecommunications subsystem would have two components, one for
receiving the data from the daughterships and one to transmit the
collected data to the Earth.  In order to receive the cross-links from
the daughterships, the mothership would carry four receivers, each of
which operates on a sub-band, allowing for simultaneous communication
with daughterships.  In order to communicate to Earth, the mothership
would use a high-gain parabolic antenna operating in Ka band, allowing
for data rates of up to~1.2~Gbps.  Handling this volume of data may
require upgrades to the current Deep Space Network (DSN) ground
stations.

At least 32 $\times$ 32~GB of data storage are required to store all
daughtership data (science data and housekeeping data).  This volume
exceeds available single chip configurations, but a RAID-0 array of
FLASH NAND storage chips would be capable of handling this volume.

In order to reduce the amount of data downlinked to the ground, it would be attractive to do the interferometric correlation on the mothership.
This solution requires very significant computing capabilities, which, in turn, require additional power.
The most likely solution is a compromise when a very limited set of sources are correlated on the mothership, but the bulk of the data downlinked to Earth to be correlated on the ground by independent research groups using own CPU capabilities.




\section{Image Reconstruction}\label{sec:image}

The act of combining the individual radio signals to form an image is an involved one for which there are several large dedicated software libraries such as AIPS and current industry standard, CASA \cite{casa07}.  However, for space based arrays, certain assumptions built into these libraries are broken.  For instance, when generating Measurement Sets (MSs), CASA takes in initial positions for the array, but assumes that the only motions of the antennas are from the fixed rotation of the Earth. Normal ground based interferometers need to be aware of all the minute effects of Earth's orbit, from daily rotation to nutation and precession, in order to get the most exact separations between antennae (called baselines) for a given target in the sky. They use and update models of the separations to add the appropriate phase delays between signals so they can be added coherently and properly correlated. For an array that is in a dynamic orbit around the Earth or Moon, there is no existing software that can simulate its observations or process its data, so we filled the gap by precomputing the baselines in the correct frame and inserting them into the CASA MS file before the visibilities are computed with standard algorithms like CASA's sm.predict().  This also allows us to utilize the full range of CASA analysis functions, including CLEAN for forming an image from the simulated data.

\subsection*{Adding Baseline Tracking}

 For space based arrays, we are only concerned with the motion of the spacecraft in the orbit, which is fully captured by an orbit file in EME2000 coordinates.  The simulated orbit determinations are made with GPS sidelobes and have an error associated with the recovered positions.  These add phase noise to the visibilities, and corrupt the recovered image.  This new noise source is included in our simulations.
 Given an RA and Declination of the source, we create a coordinate system X'Y'Z', where Z' is pointing towards the source.  RA and Dec are in the J2000 frame, which is identical to EME2000 for our purposes.

\begin{align*}
Z' & = (tx, ty, tz) \text{where} \\
tx &= cos(Dec) * cos(RA)\\
ty &= cos(Dec) * sin(RA)\\
tz &= sin(Dec)\\
\end{align*}

and RA and Dec are in radians.  

X' is formed by taking Z' x Z, which points parallel to lines of latitude and in the positive RA direction, and Y' completes the system, pointing parallel to lines of longitude and in the positive Dec direction.  

The program parses the orbit file to get the 3~dimensional difference vectors of all the pairs of daughterships over a given period.  It then projects these vector difference to the X'Y'Z' frame.  Dividing these projected differences by the user defined wavelength then gives the correct UVW coordinates for every baseline in the orbit for that particular source position.  Our codes can also select multiple time windows from the orbit file to integrate over, providing maximum flexibility for operations scheduling testing.  These coordinates are inserted into a CASA MS file after its creation, but before the radio visibilities are calculated.


\subsection*{Adding Noise}
Most radio interferometry simulation software depends on the measurement equation, where Jones matrices are used to represent the measured voltage and sky brightness for a given baseline and model the noise effects happening to it with different operations on the matrix.  Thermal noise is added directly to the brightness matrix, but for all other effects such as modeling the gain and phase errors the brightness matrix on the left and right are multiplied with the conjugate transposes of the transformation matrices.  The resulting matrix is the voltage matrix which models what the actual measurements will be.

To calculate the root mean square, RMS, on the Gaussian thermal noise to add, one usually looks at the levels of different instrumental sources of noise~\cite{Ellingson2010}.  But in this frequency range, the galactic noise is the major limiting factor.  We used Cane's 1979 measurements in~\cite{c79} to get the $T_{sys}$ temperatures from the frequency dependent galactic noise, which is turned into noise RMS using the following formula:
\begin{align*}
\sigma = \frac{2k_B T_{sys}}{\eta_s A_{eff}\sqrt{N_a(N_a - 1)N_{pol}\tau \Delta \nu}}
\end{align*}

Phase error will also be introduced in an orbiting array due to the uncertainty in the position of the daughterships, as well as differences in the clocks of the individual daughterships.  The change in phase for a given positional error $d\tau$ is $2\pi \nu d\tau$.  

From the full 3-D interferometry equation,
\begin{align*}
V_{\nu}(u, v, w) = \int \int I_{\nu}(l, m)e^{-2\pi i (ul + vm + wn)}dldm.
\end{align*}
The change in phase from a positional error $(du, dv, dw) = \frac{(dx, dy, dz)}{\lambda}$ and a clock error $dt$ is
\begin{align*}
d\phi &= 2 \pi \nu d\tau = 2 \pi \frac{c}{\lambda} d\tau = 2 \pi\left(du \cdot \ell + dv \cdot m + dw\sqrt{1 - \ell^2 - m^2} + dt\cdot \frac{c}{\lambda}\right) \\
\ell, m &<< 1 \text{ for target images of less than a few degrees} \\
\implies & \nu d\tau \approx dw + dt \cdot \nu \\ 
\implies & c d\tau \approx dz + dt\cdot c \text{, error in meters}\\ 
\implies & d\tau = \frac{dz}{c} + dt \text{, error in seconds}\\
\implies & d\phi = 2 \pi \nu d\tau = 2\pi\nu\left(\frac{dz}{c} +  dt\right) = 2\pi\left(\frac{dz}{\lambda} + \nu dt\right)
\end{align*}

Given an RMS positional uncertainty of $\sigma$, assuming uncorrelated errors, we expect the uncertainty in the z direction of a single antenna to be $\sigma_z = \frac{\sigma}{\sqrt{3}}$.  Then, since $dz$ in question is for a baseline involving 2 antennas, we get $dz$ by sampling a Gaussian distribution with $\sigma_z = \frac{\sigma\sqrt{2}}{\sqrt{3}}$.  Then clock error $dt$ is also sampled from a different Gaussian distribution with its own given RMS.  With the given uncertainties, the software can then add the phase error by applying a rotation matrix to the complex gains, or, equivalently, to the final visibilities in the CASA MS file.  The inversion of the visibilities to get the dirty image is then done with the usual 2D Inverse FFT, or can be treated with CASA's CLEAN function, which has many advanced features such as w-projection, which is useful for large images.  

\subsection*{Imaging Performance}
We were able to implement these features into an easily editable python code uploaded to https://github.com/alexhege/SpaceCASA .  The following figures were created with this script to emulate the performance of the 32 element RELIC array.  We tested an average DRAGN for the array, 100 arcsec wide, 100 Jy total brightness, and used a detailed picture of Cygnus A at 21 cm to provide realistic complexity to the truth image.  We are integrating for 35 minutes, which is 2100 seconds, again a typical expectation.  We assume a 5 nanosecond phase uncertainty per spacecraft, and galactic noise at 10 MHz.  We are using CASA's CLEAN algorithm to create the image and have w-projection turned on.  We have shrunk the baselines by a factor of 10, which gives us an image 10x larger than our prescribed 100 arcsec.  This was done since CASA crashes if the baselines are too large.  But the relative scales are the same, so it provides an accurate sense of RELIC's performance.  

\begin{figure}[h]
    \begin{center}
              \subfigure[Input image of realistic DRAGN, actually Cygnus A at 21 cm wavelength, scaled to 100 arcsec.] {\includegraphics[width=.45\textwidth]{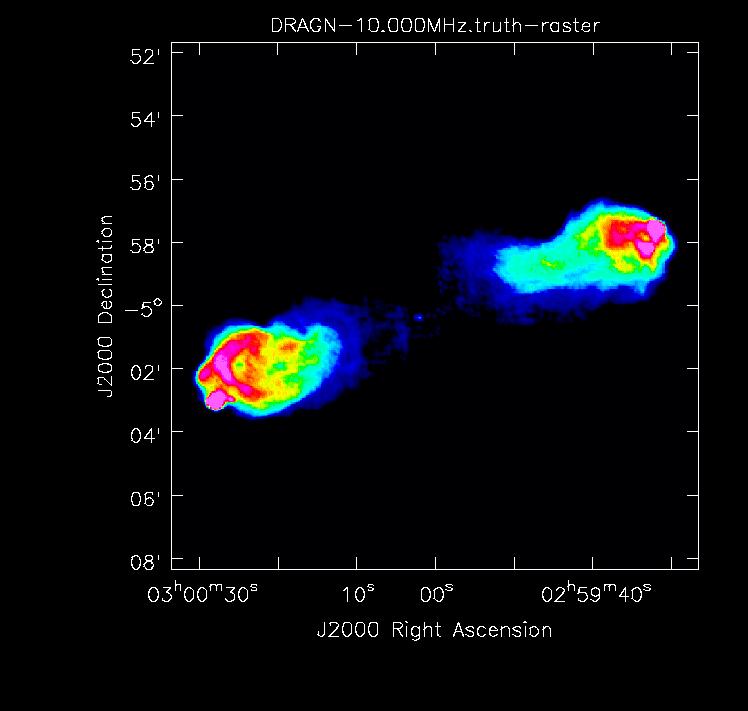}}
              \subfigure[Synthesized beam from point spread function after 35 minutes of integration] {\includegraphics[width=.45\textwidth, height=.43\textwidth]{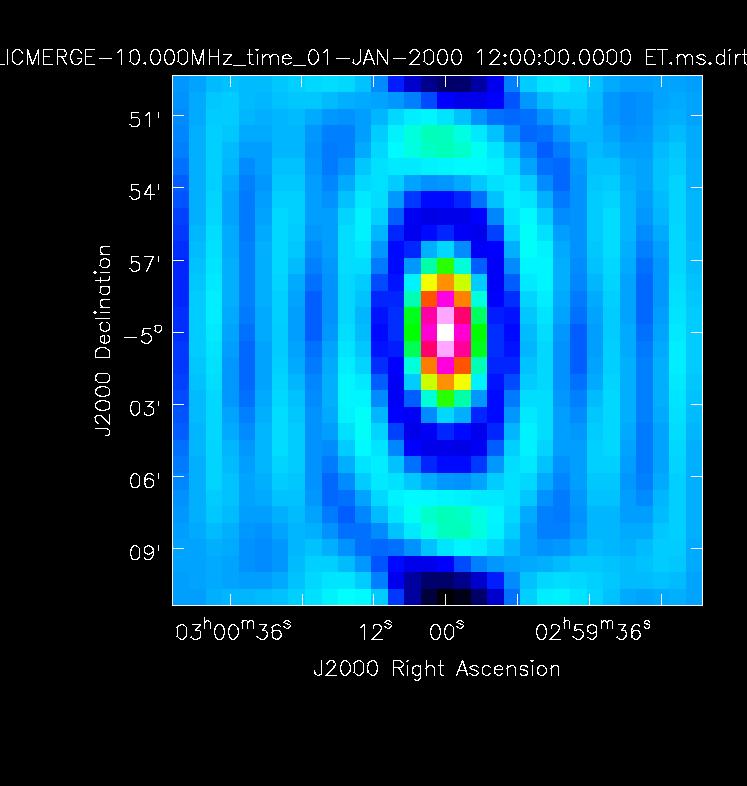}}
              \subfigure[Smoothed recovered image after 35 minutes of integration] {\includegraphics[width=.45\textwidth]{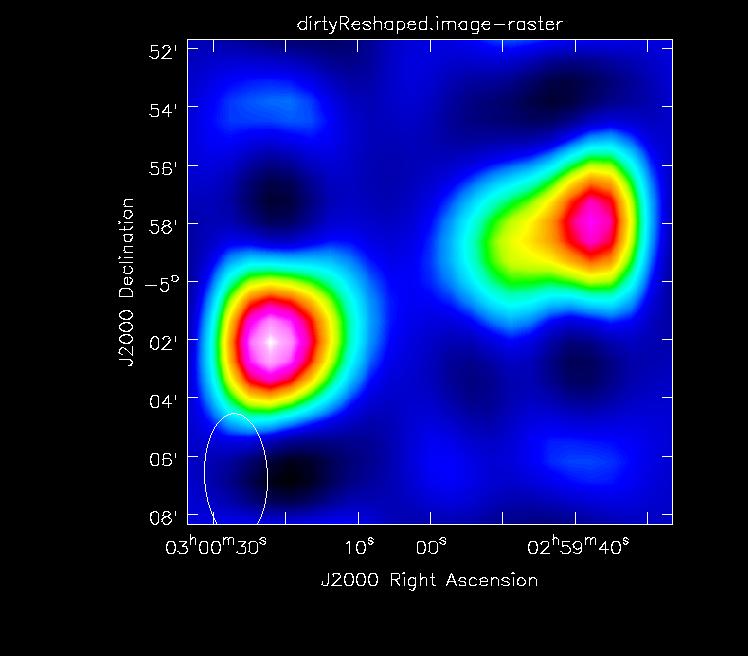}}
              \subfigure[Root Mean Square Error between Truth and Recovered Image] {\includegraphics[width=.45\textwidth]{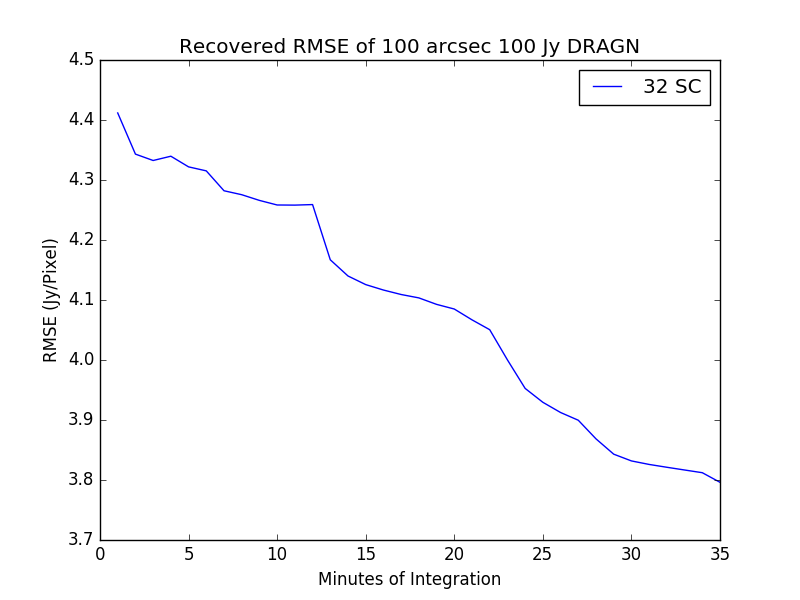}}
        \caption{Images showing the CASA simulated performance of RELIC on a 100 arcsecond wide, 100 Jy total bright DRAGN.  This was done with part of the orbit where the maximum baseline is ~370 km, which corresponds to a resolution of roughly 15-20 arcsec.}
    \end{center}
\label{fig:casaimage} 
\end{figure}

\section{Discussion}\label{sec:conclude}

We described a proposed space-based decametric wavelength radio telescope aimed primarily for observation of radio galaxies.  It leverages on existing small satellite technologies with the baseline mission consisting of one mothership and 32 daugterships with dual polarization antennas. 
Based on our analysis of the uv-plane coverage provided Section~\ref{sec:mission.orbits}, the necessary integration times discussed at the beginning of Section~\ref{sec:science} and the data downlink requirements discussed in Sections~\ref{sec:mission.comm} and~\ref{datarate:subsec}, the proposed mission can fit under a standard two-year mission plan to fulfill the proposed scientific goals.
The proposed space-based telescope being free from the influence if the Earth's ionosphere will become the key part of a multi-wavelength study of physical processes in the active galaxies and will become the primary tool for observing low-frequency phenomena from wide range of radio sources in the sky.
The proposed mission is viable for implementation in the near term based missions manifested for launch in the next 5 years.

\begin{acknowledgments}
We thank S.~Murray and K.~Weiler for their guidance and helpful comments at an early stage of this concept development as well as R.~MacDowall for illuminating conversations.
Some of the science motivation for the RELIC concept was articulated
in ``Small Satellites: A Revolution in Space Science'' study co-led by
Charles Norton, Sergio Pellegrino, and Michael Johnson at the W.~M.~Keck Institute for Space Studies.
Part of this research was carried
out at the Jet Propulsion Laboratory, California Institute of
Technology, under a contract with the National Aeronautics and Space Administration.
Copyright 2017. All rights reserved.
\end{acknowledgments}

\bibliographystyle{unsrt}

\begin{thebibliography}{100}

\bibitem[\protect\citeauthoryear{Jansky}{1935}]{j35} Jansky, K.~G.  1935, 
        Proc.\ I.\ R.\ E., 23, 1158 

\bibitem[\protect\citeauthoryear{Ryle \& Vonberg}{1946}]{rv46} Ryle,
        M.\ \& Vonberg, D.~D.  1946, \nat, 158, 339

\bibitem[\protect\citeauthoryear{Pawsey, Payne-Scott, \&
        McCready}{Pawsey et al.}{1946}]{pp-sm46} Pawsey, J.~L.,
        Payne-Scott, R., \& McCready, L.~L.  1946, \nat, 157, 158

\bibitem[\protect\citeauthoryear{McCready, Pawsey, \&
        Payne-Scott}{McCready et al.}{1947}]{mpp-s47} McCready,
        L.~L,. Pawsey, J.~L., \& Payne-Scott, R.  1947, Proc.\ Royal
        Soc.~A, 190, 357

\bibitem[\protect\citeauthoryear{Ryle, Smith, \& Elsmore}{Ryle et
        al.}{1950}]{rse50} Ryle, M., Smith, F.~G., \& Elsemore, B.
        1950, \mnras, 110, 508

\bibitem[\protect\citeauthoryear{Mills}{1952}]{m52} Mills, B.~Y.
        1952, Aust.\ J.\ Sci.\ Res., A5, 456

\bibitem[\protect\citeauthoryear{Ryle}{1952}]{r52} Ryle, M.  1952,
        Proc.\ Royal Soc.~A, 211, 351
        
\bibitem[\protect\citeauthoryear{Reber \& Ellis}{1956}]{re56}
	Reber, G., \& Ellis, G.~R.  1956, 
	``Cosmic Radio-Frequency Radiation Near One Megacycle,''
	\jgr, 61, 1

\bibitem[\protect\citeauthoryear{Cane}{1979}]{c79}
        Cane, H.~V.  1979,
        ``Spectra of the Non-Thermal Radio Radiation from the Galactic
        Polar Regions,''
        \mnras, 189, 465	
	
\bibitem[\protect\citeauthoryear{Ellis \& Mendillo}{1987}]{em87}
	Ellis, G.~R.~A., \& Mendillo, M.  1987,
	``A 1.6~MHz survey of the galactic background radio emission,''
	Aust.\ J.\ Phys., 40, 705

\bibitem[\protect\citeauthoryear{Salas et al.}{2017}]{sovw+17}
	Salas, P., Oonk, J.~B.~R., van~Weeren, R.~J., Salgado, F.,
	Morabito, L.~K., Toribio, M.~C., Emig, K., Rottgering,
	H.~J.~A., \& Tielens, A.~G.~G.~M.  2017,
	\mnras, in press
	
\bibitem[\protect\citeauthoryear{Handbook Radio Astronomy}{2004}]{hra2004}	
	Handbook Radio Astronomy, Second Edition, International Telecommunications Union, Geneva, 2004.

\bibitem[\protect\citeauthoryear{Bougeret et al.}{1995}]{wind}
	Bougeret, J.-L., Kaiser, M.~L., Kellogg, P.~J., et al.  1995,
	``Waves: The Radio and Plasma Wave Investigation on the Wind Spacecraft,''
	Space Sci.\ Rev., 71, 231; doi: 10.1007/BF00751331

\bibitem[\protect\citeauthoryear{Gurnett et al.}{2004}]{cassini}
	Gurnett, D.~A., Kurth, W.~S., Kirchner, D.~L., et al.  2004,
	``The Cassini Radio and Plasma Wave Investigation,''
	Space Sci.\ Rev., 114, 395; doi: 10.1007/s11214-004-1434-0

\bibitem[\protect\citeauthoryear{Bougeret et al.}{2008}]{stereo}
	Bougeret, J.~L., Goetz, K., Kaiser, M.~L., et al.  2008,
	``\hbox{S/WAVES}: The Radio and Plasma Wave Investigation on the STEREO Mission,''
	Space Sci.\ Rev., 136, 487; doi: 10.1007/s11214-007-9298-8

\bibitem[\protect\citeauthoryear{Kurth et al.}{2012}]{juno}
	Kurth, W.~S., Kirchner, D.~L., Hospodarsky, G.~B., et al.  2012,
	``The Juno Waves investigation,''
	European Planetary Science Congress 2012, id.~EPSC2012-281; http://meetings.copernicus.org/epsc2012

\bibitem[\protect\citeauthoryear{Alexander et al.}{1969}]{abcsw69}
        Alexander, J.~K., Brown, L.~W., Clark, T.~A., Stone, R.~G., \&
        Weber, R.~R.  1969,
        ``The Spectrum of the Cosmic Radio Background Between~0.4 and~6.5~MHz,''
        \apj, 157, L163

\bibitem[\protect\citeauthoryear{Alexander et al.}{1975}]{akngw75}
        Alexander, J.~K., Kaiser, M.~L., Novaco, J.~C., Grena, F.~R.,
        \& Weber, R.~R.  1975,
        ``Scientific instrumentation of the Radio-Astronomy-Explorer-2
        satellite,''
        \aap, 40, 365

\bibitem[\protect\citeauthoryear{Swarup}{1990}]{s90}
        Swarup, G.  1990,
        ``Giant metrewave radio telescope (GMRT) - Scientific objectives and design aspects,''
        Indian J.\ Radio Space, 19, 493

\bibitem[\protect\citeauthoryear{Ananthakrishnan}{1995}]{a95}
        Ananthakrishnan, S.  1995,
        ``The Giant Meterwave Radio Telescope / \hbox{GMRT},''
        J.\ Astrophys.\ Astron., 16, 427

\bibitem[\protect\citeauthoryear{Kassim et al.}{2007}]{vla74}
        Kassim, N.~E., Lazio, T.~J.~W., Erickson, W.~C., Perley,
        R.~A., Cotton, W.~D., Greisen, E.~W., Cohen, A.~S., Hicks, B.,
        Schmitt, H.~R., \& Katz, D. 2007,
        ``The 74~MHz System on the Very Large Array,''
        \apjs, 172, 686
        
\bibitem[\protect\citeauthoryear{Braude et al.}{1978}]{bmrsz78}
        Braude, S.~Ia., Megn, A.~V., Riabov, B.~P., Sharykin, N.~K.,
        \& Zhuk, I.~N.  1978,
        ``Decametric survey of discrete sources in the Northern sky. I~-~The UTR-2 radio telescope: Experimental techniques and data processing,''
        \apss, 54, 3
        
\bibitem[\protect\citeauthoryear{van~Haarlem et al.}{2013}]{lofar}
	van Haarlem, M.~P., et al.  2013, 
	``\hbox{LOFAR}: The LOw-Frequency ARray,''
	\aap, 556, A2

\bibitem[\protect\citeauthoryear{Ellingson et al.}{2013}]{lwa1}
        Ellingson, S.~W., Taylor, G.~B., Craig, J., et al.  2013,
        ``The LWA1 Radio Telescope,''
        IEEE Trans.\ Ant.\ Prop., 61, 2540; doi: 10.1109/TAP.2013.2242826

\bibitem[\protect\citeauthoryear{Taylor et al.}{2012}]{tek+12}
        Taylor, G.~B., Ellingson, S.~W., Kassim, N.~E., et al.  2012, 
        ``First Light for the First Station of the Long Wavelength Array,''
        J.\ Astron.\ Instrum., 1, 1250004; doi: 10.1142/S2251171712500043

\bibitem[\protect\citeauthoryear{Tingay et al.}{2013}]{mwa}
	Tingay, S.~J., et al.  2013, 
	``The Murchison Widefield Array: The Square Kilometre Array Precursor at Low Radio Frequencies,''
	Publ.\ Astron.\ Soc.\ A., 30, 7

\bibitem[\protect\citeauthoryear{French et al.}{1967}]{fhr67}
	French, F.~W., Huguenin, G.~R., \& Rodman, A.~K.  1967,
	``A synthetic aperture approach to space-based radio telescopes,''
	J.\ Spacecraft Rockets, 4, 1649

\bibitem[\protect\citeauthoryear{Weiler et al.}{1988}]{wjs+88}
	Weiler, K.~W., Johnston, K.~J., Simon, R.~S., Dennison, B.~K.,
	Erickson, W.~C., Kaiser, M.~L., Cane, H.~V., \& Desch, M.~D.
	1988, 
	\aap, 195, 372

\bibitem[\protect\citeauthoryear{Basart et al.}{1997a}]{bbd+97a}
	Basart, J.~P., Burns, J.~O., Dennison, B.~K., Weiler, K.~W.,
Kassim, N.~E., Castillo, S.~P., \& McCune, B.~M.  1997a,
	``Directions for space-based low frequency radio astronomy.\ 1.~System considerations,''
	Radio Sci., 32, 251

\bibitem[\protect\citeauthoryear{Basart et al.}{1997b}]{bbd+97b}
	Basart, J.~P., Burns, J.~O., Dennison, B.~K., Weiler, K.~W.,
Kassim, N.~E., Castillo, S.~P., \& McCune, B.~M.  1997b,
	``Directions for Space-Based Low-Frequency Radio Astronomy 2.~Telescopes,''
	Radio Sci., 32, 265

\bibitem[\protect\citeauthoryear{Oberoi \& Pin{\c{c}}on}{2005}]{op05}
	Oberoi, D., \& Pin{\c{c}}on, J.-L.  2005,
	Radio Sci., 40, 4004
   
\bibitem[\protect\citeauthoryear{Jones et al.}{2000}]{alfa}
	Jones, D.~L., et al.  2000,
	``The ALFA Medium Explorer Mission,''
	Adv.\ Space Res., 26, 743

\bibitem[\protect\citeauthoryear{Banazadeh et al.}{2013}]{bljsfa13}
	Banazadeh, P., Lazio, J., Jones, D., Scharf, D.~P., Fowler,
	W., \& Aladangady, C.  2013,
	``Feasibility analysis of \hbox{XSOLANTRA}: A mission concept to detect exoplanets with an array of CubeSats,''
	in Proc.\ 2013 IEEE Aerospace Conf.; 
	doi: 10.1109/AERO.2013.6496864
	
	

\bibitem[\protect\citeauthoryear{Rajan et al.}{2011}]{Rajan2011}
	R. T. Rajan, S. Engelen, M. Bentum and C. Verhoeven, 2011,
	``Orbiting Low Frequency Array for radio astronomy,''
	in Proc. 2011 Aerospace Conference, 1-11, doi=10.1109/AERO.2011.5747222
	
doi: 10.1109/AERO.2016.7500678
\bibitem[\protect\citeauthoryear{A. J. Boonstra et al}{2016}]{Boonstra2016}
	A. J. Boonstra et al, ``Discovering the sky at the Longest Wavelengths (DSL),''
	in Proc. 2016 IEEE Aerospace Conference, Big Sky, MT, 2016, pp. 1-20.
	

\bibitem[\protect\citeauthoryear{Cecconi et al.}{2018}]{noire2018}
	     B.~Cecconi et al.,
	     "NOIRE Study Report: Towards a Low Frequency Radio
	     Interferometer in Space," In Proc. 2018 IEEE Aerospace Conference, arXiv:1710.10245

\bibitem[\protect\citeauthoryear{Baumback et al.}{1986}]{bgcs86}
        Baumback, M.~M., Gurnett, D.~A., Calvert, W., \& Shawhan,
        S.~D.  1986,
	``Satellite interferometric measurements of auroral kilometric radiation,''
	Geophys.\ Res.\ Lett., 13, 1105

\bibitem[\protect\citeauthoryear{Mutel et al.}{2004}]{mgc04}
        Mutel, R., Gurnett, D.~A., \& Christopher, I. 2004,
	``Spatial and Temporal Properties of AKR Burst Emission Derived From Cluster WBD VLBI Studies,''
	Annales Geophys., 22, 2625

\bibitem[\protect\citeauthoryear{Norton et al.}{2014}]{kiss}
	Norton, C.~D., Pellegrino, S., Johnson, M., et al.  2014,
	``Small Satellites: A Revolution in Space Science,''
	W.~M.~Keck Institute for Space Studies; http://kiss.caltech.edu/study/smallsat/KISS-SmallSat-FinalReport.pdf
	
\bibitem[\protect\citeauthoryear{Kassim \& Weiler}{1990}]{kw90}
	Kassim, N.~E., \& Weiler, K.~W.  1990,
	\textit{Low frequency astrophysics from space}, Lecture Notes
	in Physics, Vol.~362 (Springer-Verlag: Berlin)

\bibitem[\protect\citeauthoryear{Stone et al.}{2000}]{swgb00}
	Stone, R.~G., Weiler, K.~W., Goldstein, M.~L., \& Bougeret,
	J.-L.  1998, 
	\textit{Radio Astronomy at Long Wavelengths}, Geophysical
Monograph Series, Vol.~119 (American Geophysical Union: Washington, DC) ISSN 0065-8448
	
\bibitem[\protect\citeauthoryear{Harris}{2005}]{h05}
	Harris, D.~E.  2005,
	``From Clark Lake to Chandra: Closing in on the Low End of the Relativistic Electron Spectra in Extragalactic Sources,''	
	in \textit{From Clark Lake to the Long Wavelength Array: Bill
Erickson's Radio Science}, Astron.\ Soc.\ Pacific Conference Series,
Vol.~345, eds.\ N.~Kassim, M.~Perez, M.~Junor, \& P.~Henning (Astron.\
Soc.\ Pacific: San Francisco) p.~254

\bibitem[\protect\citeauthoryear{Ineson et al.}{2017}]{ichm17}
	Ineson, J., Croston, J.~H., Hardcastle, M.~J., \& Mingo, B.  2017,
	``A representative survey of the dynamics and energetics of FRII radio galaxies,''
	\mnras, in press

\bibitem{Laing1983} R.~Laing, J.~Riley and M.~Longair, {\em Bright radio sources at 178 MHz: Flux densities, optical identifications and the cosmological evolution of powerful radio galaxies}, Mon. Not. R. astr. Soc., {\bf204}, 151-187 (1983).

\bibitem{McKean2016}J. P. ~McKean et al., {\em LOFAR imaging of Cygnus A ? direct detection of a turnover in the hotspot radio spectra}, Mon. Not. R. astr. Soc., {\bf463}, 3, 3143?3150 (2016).




\bibitem{KleinWolt2012}M.~Klein Wolt et al., {\em Radio astronomy with the European Lunar Lander: Opening up the last unexplored frequency regime,} Planetary and Space Science, {\bf 74}, 1, 167-178 (2012). 

\bibitem{Chen2018}L.~Chen et al., {\em Antenna design and implementation for the future space Ultra-Long wavelength radio telescope}, submitted to Experimental Astronomy, arXiv:1802.07640

\bibitem{Cordes1990} J.M.~Cordes, {\em Low frequency interstellar scattering and pulsar observations}, in Low frequency astrophysics from space; Proceedings of an International Workshop, Crystal City, VA, Jan. 8, 9, 1990 (A91-57026 24-89). Berlin and New York, Springer-Verlag, 165-174 (1990).

\bibitem{Braude:2002}S.Ya.~Braude et al, {\em Decametric survey of discrete sources in the northern sky}, Astrophys and Sp. Sci., {\bf 280} 3, 235--300 (2002). 

\bibitem{Chien2000_6} S.~Chien, G.~Rabideau, R.~Knight, R.~Sherwood, B.~Engelhardt, D.~Mutz, T.~Estlin, B.~Smith, F.~Fisher, T.~Barrett, G.~Stebbins, Aspen?automated planning and scheduling for space mission operations. InSpace Ops 2000 June, Toulouse, France, AIAA.

\bibitem{Chien2000_4} S.A.~Chien, R.~Knight, A.~Stechert, R.~Sherwood, G.~Rabideau, ``Using Iterative Repair to Improve the Responsiveness of Planning and Scheduling,'' In Artificial Intelligence Planning Systems, 2000 Apr (pp. 300-307), AAAI Press.

\bibitem{Acton1996} C.H.~Acton, ``Ancillary Data Services of NASA's Navigation and Ancillary Information Facility,'' Planetary and Space Science, Vol. 44, No. 1, pp. 65-70, 1996.

\bibitem[\protect\citeauthoryear{Mills}{1953}]{m53}
	Mills, B.~Y.  1953,
	``The Radio Brightness Distributions over Four Discrete Sources of Cosmic Noise,''
	Aust.\ J.\ Phys., 6, 452

\bibitem[\protect\citeauthoryear{Blythe}{1957}]{b57}
	Blythe, J.~H.  1957,
	``A new type of pencil beam aerial for radio astronomy,''
	\mnras, 117, 644

\bibitem[\protect\citeauthoryear{Ryle \& Hewish}{1960}]{rh60}
	Ryle, M., \& Hewish, A.  1960,
	``The synthesis of large radio telescopes,''
	\mnras, 120, 220


\bibitem[\protect\citeauthoryear{Bell et al.}{2014}]{bskk14}
Bell, D., Satorius, E., Kuperman, I., \& Koenig, J.  2014,
``Multiuser Receiver Architectures for Space Modems,''
The Interplanetary Network Progress Report, vol.~42-198 (Jet Propulsion Laboratory, California Institute of Technology: Pasadena, CA) p.~1-13

\bibitem{porat} B.~Porat, {\em A Course in Digital Signal Processing}, John Wiley and Sons, New York, 1997.

\bibitem{DSN} {\em Pseudo-Noise and Regeenerative Ranging}, Deep Space Network No. 810-005 214, Rev A, October 28, 2015

\bibitem{Berner} J. Berner, S. Bryant, P Kinman, {\em Range Measurement as Practiced in the Deep Space Network}, Proceedings of the IEEE, Vol. 95, No. 11, November 2007.

\bibitem{TLMRNG} K. Andrews J. Hamkins, S. Shambayati, V. Vilnrotter, {\em Telemetry-Based Ranging}, Proceedings of the 2010 IEEE Aerospace Conference, Big Sky, MT, March 2010.

\bibitem[\protect\citeauthoryear{Thompson, Moran, \& Swenson}{2007}]{tms07}
	Thompson, A.~R., Moran, J.~M., \& Swenson, G.~W.  2007,
	\textit{Interferometry and Synthesis in Radio Astronomy},
	(Wiley: New York)


\bibitem[\protect\citeauthoryear{McMullin et al.}{2007}]{casa07} McMullin, J. P., Waters, B., Schiebel, D., Young, W., and Golap, K. 2007, Astronomical Data Analysis Software and Systems XVI (ASP Conf. Ser. 376), ed. R. A. Shaw, F. Hill, D. J. Bell (San Francisco, CA: ASP), 127

\bibitem[\protect\citeauthoryear{Kocz et al.}{2015}]{kgb+15}
	Kocz, J., Greenhill, L.~J., Barsdell, B.~R., Price, D.,
	Bernardi, G., Bourke, S., Clark, M.~A., Craig, J., Dexter, M.,
	Dowell, J., Eftekhari, T., Ellingson, S., Hallinan, G.,
	Hartman, J., Jameson, A., MacMahon, D., Taylor, G., Schinzel, F., \&
	Werthimer, D.  2015, ``Digital Signal Processing Using Stream
	High Performance Computing: A 512-Input Broadband Correlator
	for Radio Astronomy,'' J.\ Astron.\ Instrumentation, 4,
	1550003; doi: 10.1142/S2251171715500038

\bibitem{APSYNSIM} I.~Marti-Vidal, {\em Aperture Synthesis Simulator for Radio Astronomy}, https://launchpad.net/apsynsim

\bibitem{Ellingson2010} S.~Ellingson, {\em Sensitivity of Antenna Arrays for Long-Wavelength Radio Astronomy}, IEEE Transactions on Antennas and Propagation (2010).

\bibitem[\protect\citeauthoryear{Zurbuchen et al.}{2016}]{cubesats}
	Zurbuchen, T.~H., Bhavya Lal, B., et al.  2016, \textit{Achieving Science with CubeSats: Thinking Inside the Box}
	(National Academies Press: Washington, DC) ISBN: 978-0-309-44263-3
	
\bibitem{Meyer-Vernet:1989} N.Meyer-Vernet and C.Perche {\em Tool Kit for Antennae and Thermal Noise Near the Plasma Frequency}, Journal of Geophysical Research, 94, 2405 ? 2415, (1989).

\bibitem{Zaslavsky:2011} A.Zaslavsky et al.,  {\em On the antenna calibration of space radio instruments using the galactic background: General formulas and application to STEREO/WAVES}, Radio Science, 46, RS2008 (2011).

\bibitem{James:2015} James, H. G., King, E.P., White, A., Hum, R. H., Lunscher, W. H. H. L., Siefring, C. L., {\em The e-POP Radio Receiver Instrument on CASSIOPE}, Space Sci Rev, 189, 79-105 (2015).

\end{thebibliography}

\newpage{\pagestyle{empty}\cleardoublepage}

\appendix
\section{RELIC HF Antenna and Front-End Design}

While the detailed description of the science data acquisition hardware is beyond the scope of this paper, it is nonetheless an integral part of the mission concept description and it is also responsible for generating the raw science data that are then transmitted by the daughter ships (which is much of the focus of this paper).  Accordingly, we present here a brief description of how the science data acquisition hardware for RELIC might look.

\subsection{Objectives}

The objective of this is to characterize the antenna length and front-end impedance parameters that provide a background-noise-limited system for the RELIC antenna. The RELIC signals will be wide-band measurements across the 0.1-30 Mhz band. The overall desired performance is 1) that the antenna noise be greater than the amplifier noise, 2) stable gain in each sub-band, 3) smooth gain and phase variations across the band, 4) reducing too much variation of gain across the band so that dynamic range in not an issue. The scope of this document is to identify a solution for a high-impedance input design that primarily aims to meet point 1 but with points 2-4 in mind. 


\subsection{Assumptions}
\subsubsection{Antenna Model}

\begin{figure}[htb]
\includegraphics[width=0.75\textwidth]{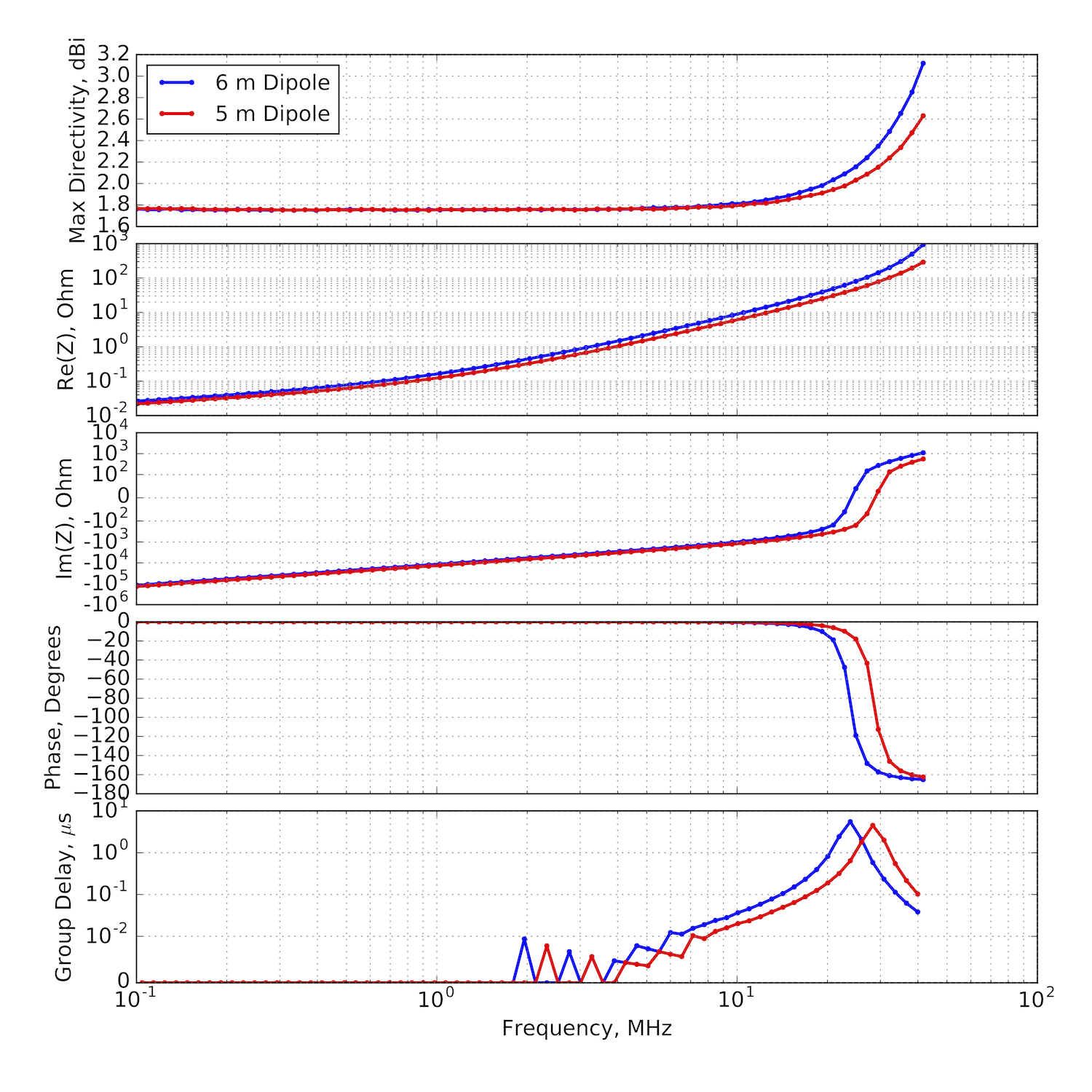}
\caption{NEC2 models of a 6-meter and 5-meter full-length dipole. The model assumes the antenna is made of copper rod with 0.6 cm diameter floating in free space.}
\label{ant_models:fig}
\end{figure}

The band of interest for RELIC extends from 0.1 -- 30 MHz. Below 1 MHz, the galactic noise is low but ``shot noise'' due to the plasma environment is dominant. The antenna models used for the initial assessment are 6-meter and 5-meter full length dipoles made of 0.6 cm ($\sim$0.25'') diameter copper. The antennas are modeled using NEC2. The directivity, impedance (real and imaginary) along with the phase and group delay are shown in Figure~\ref{ant_models:fig}.

\subsubsection{Background Noise}

\begin{figure}[htb]
\includegraphics[width=0.5\textwidth]{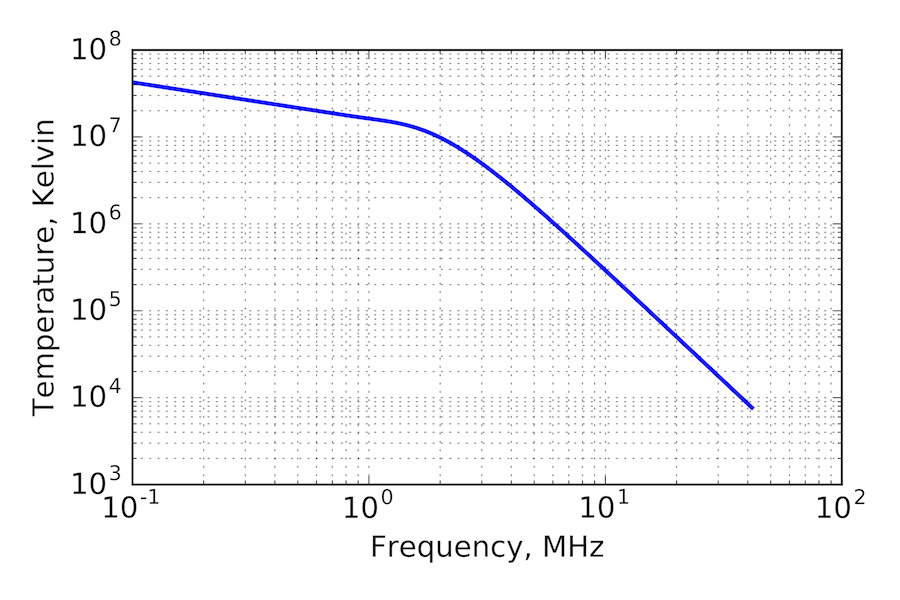}
\caption{The galactic noise temperature from the parameterization in~\cite{c79}.}
\label{Gal_Temperature:fig}
\end{figure}

The galactic noise is modeled according to the parametrization described in \cite{c79}. A plot of the average galactic noise temperature  is shown in Figure~\ref{Gal_Temperature:fig} 

\begin{figure}[htb]
\includegraphics[width=0.5\textwidth]{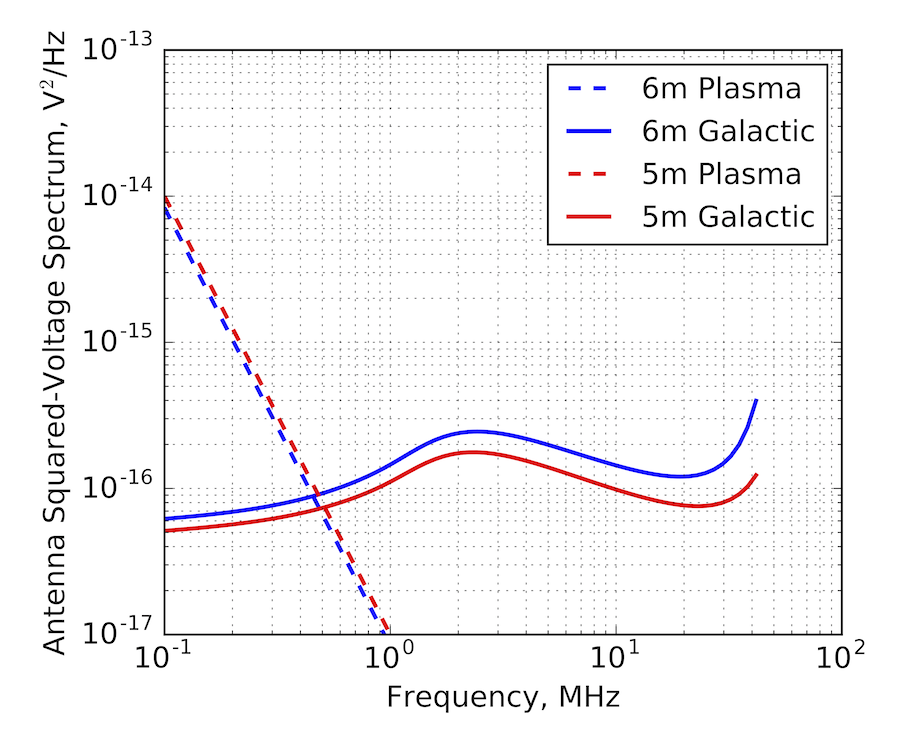}
\caption{The noise-squared voltage at the antenna terminals for the plasma shot noise contribution and the galactic noise. The noise is modeled for a 6 m and 5m full-length dipole.}
\label{ant_noise:fig}
\end{figure}

The noise-voltage-squared spectrum at the antenna terminals due to galactic noise is given by $V^2_{A,gal}=4k_BT_{gal}R_{ant}$, where $k_B$ is Boltzmann's constant and $R_{ant}$is the real part of the antenna impedance. In addition to galactic noise, there is a ``shot noise'' contribution due to electrons in the plasma colliding with the antenna and inducing currents~\cite{Meyer-Vernet:1989}. The noise-voltage-squared contribution as parameterized by~\cite{Zaslavsky:2011} is given by:
\begin{equation}
V^2_{A,QTN}=5\times10^{-5}\frac{V^2}{Hz}\times\left(\frac{n_e}{cm^{-3}}\right)\times\left(\frac{T_e}{K}\right)\times\left(\frac{f}{\text{Hz}}\right)^3\times\left(\frac{L/2}{\text{m}}\right)^{-1},
\end{equation}
where $n_e$ is the plasma frequency in electrons per cm$^3$, $T_e$ is the electron plasma temperature in Kelvin, $f$ is the frequency in Hz, and $L$ is the full dipole length in meters. We assume $n_e$=5~cm$^{-3}$ and $T_e$=10$^5$ Kelvin. The background noise-voltage-squared contributions are shown for a modeled 6-meter (full length) and a 5-meter dipole made of 0.6 cm diameter copper rod in Figure~\ref{ant_noise:fig}.

\subsection{Analysis}

The goal is to estimate the sources of noise at the first amplifier in the signal chain to ensure the noise is dominated by external backgrounds rather than by the amplifier itself.  The external noise sources producing voltages at the antenna terminals with produce a voltage at the load according to

\begin{equation}
V_L=\frac{Z_L}{Z_A+Z_L}V_A,
\end{equation}

where $Z_L$ is the complex impedance of the load and $Z_A$ is the antenna impedance. The noise due to the amplifier, at the load is given by $V^2_{L,amp}=k_BT_{amp}R_{amp}$. We consider an operation amplifier (OpAmp) approach for producing a high-input impedance to the amplifier as seen from the terminals of the antenna.

\begin{figure}[htb]
\includegraphics[width=0.70\textwidth]{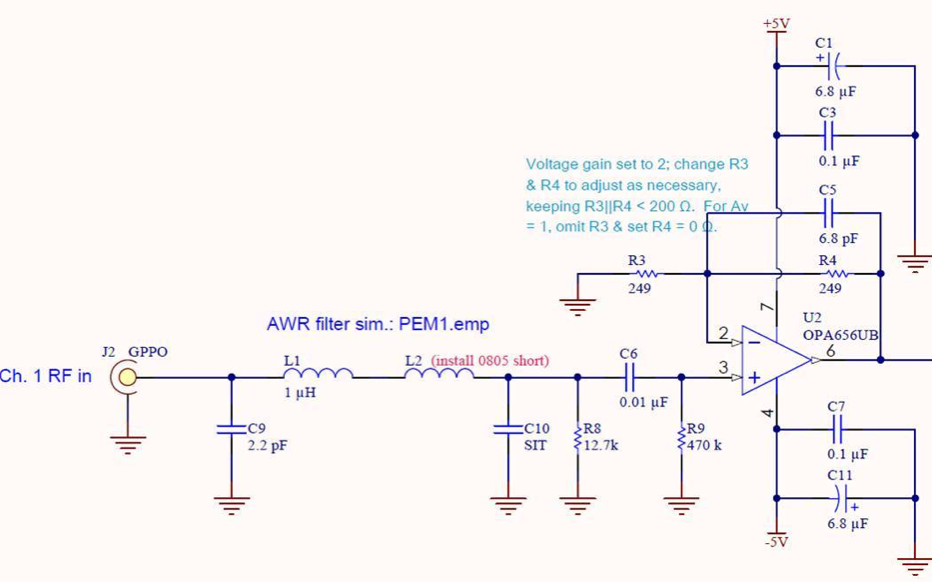}
\caption{Operational Amplifier front end design by Bob Dengler for the Universal Space Transponder.}
\label{OpAmpFrontEnd:fig}
\end{figure}

OpAmps can be designed to have a high input impedance but modeling their noise is somewhat more involved. A modeling effort using an OpAmp was done for JPL?s Universal Space Transponder (UST) Jovian burst science application development. Figure~\ref{OpAmpFrontEnd:fig} is an OPA656-based design by Robert Dengler for the Universal Space Transponder (UST) using 12.7 k$\Omega$ input impedance (set by R8 in Figure~\ref{OpAmpFrontEnd:fig}). 

For this application, we considered the OPA656 from Texas Instruments. The choice is driven by the noise characteristics $V_n=7\text{nV}/\sqrt{Hz}$, $I_n=1.3\text{fA}/\sqrt{Hz}$. The low current noise is particularly important for these high impedance applications. It is worth mentioning the e-POP radio receiver instrument on CASSIOPE also used OPA656 for its receiver with a 100M$\Omega$ input impedance~\cite{James:2015}. 

\begin{figure}[h]
 \centering
 \subfigure[$R_8=$12.8k$\Omega$]{%
	\includegraphics[width=.3\textwidth]{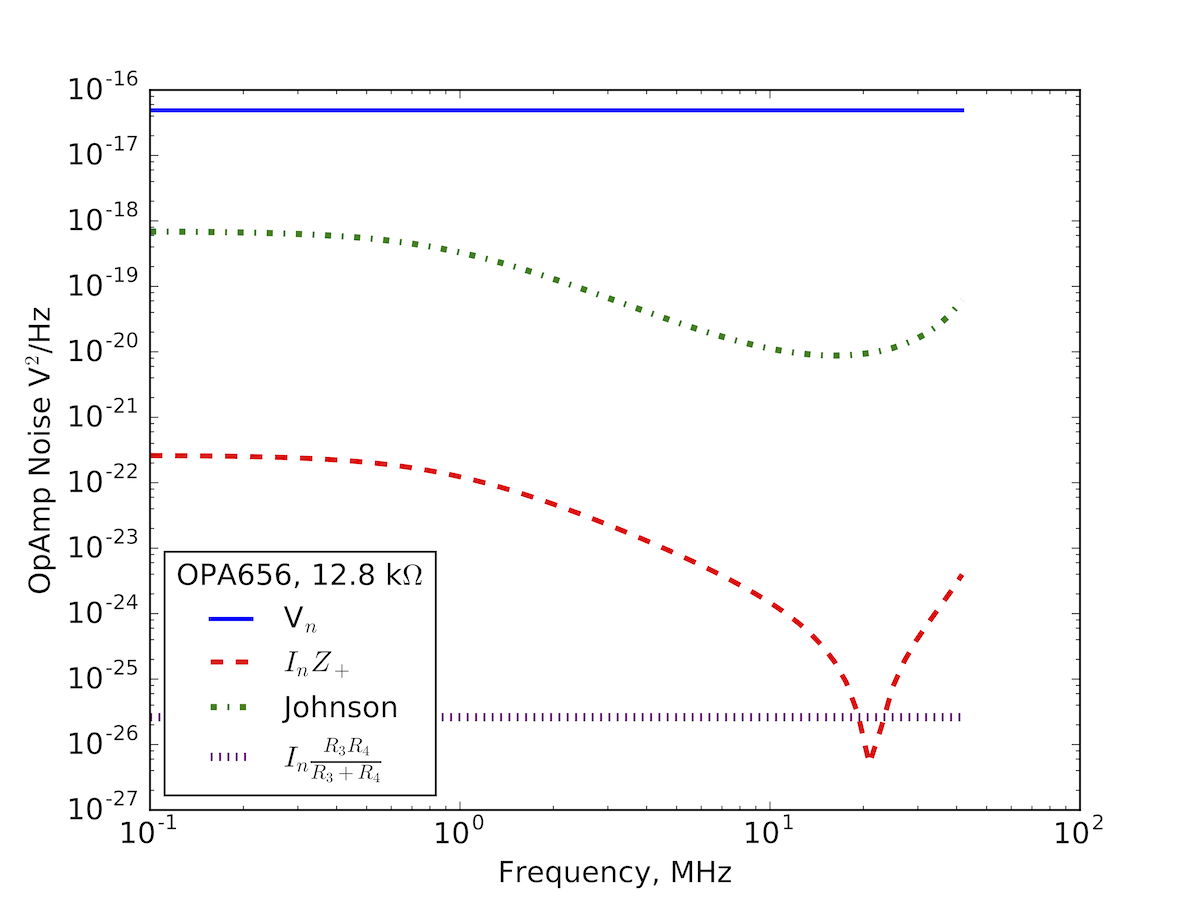}
		}
\subfigure[$R_8=$50k$\Omega$]{%
	\includegraphics[width=.3\textwidth]{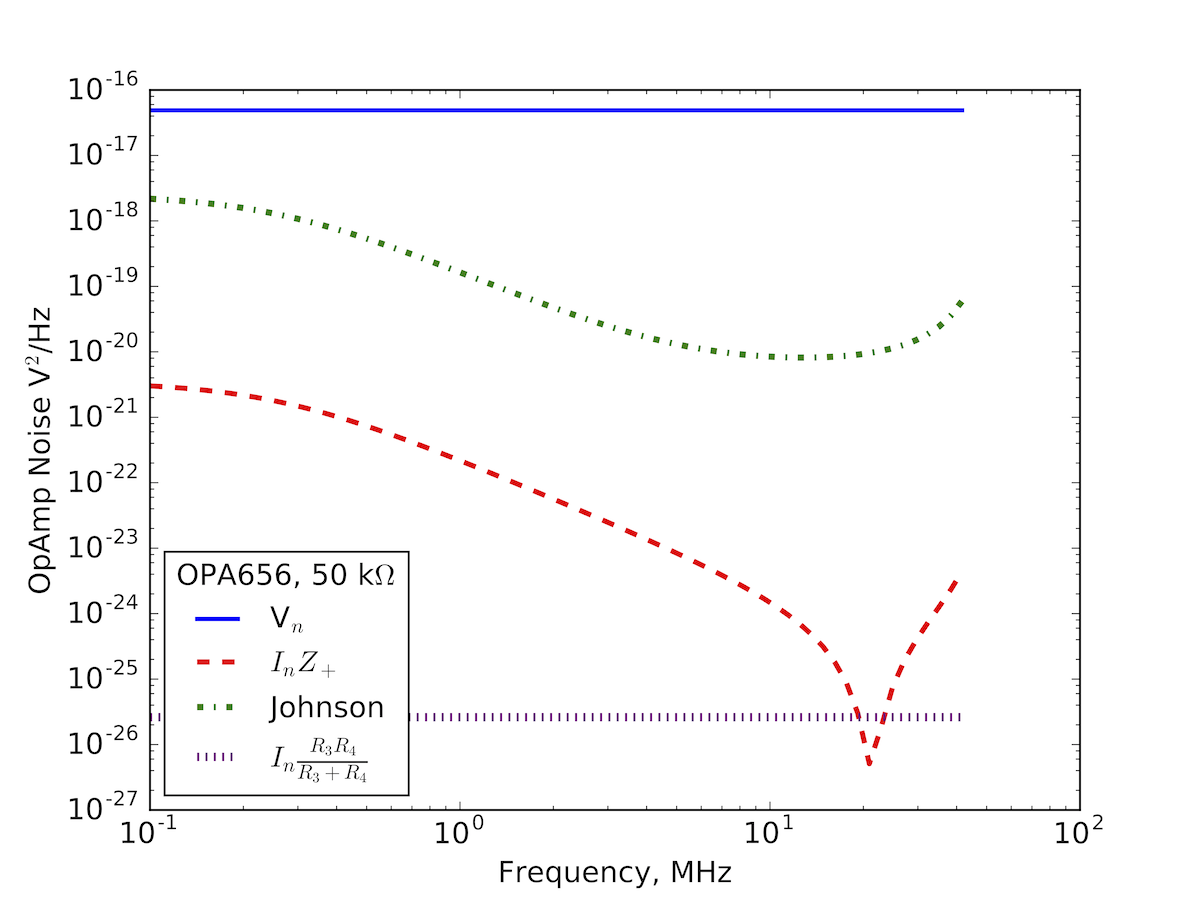}
		}
 \subfigure[$R_8=$100k$\Omega$]{%
	\includegraphics[width=.3\textwidth]{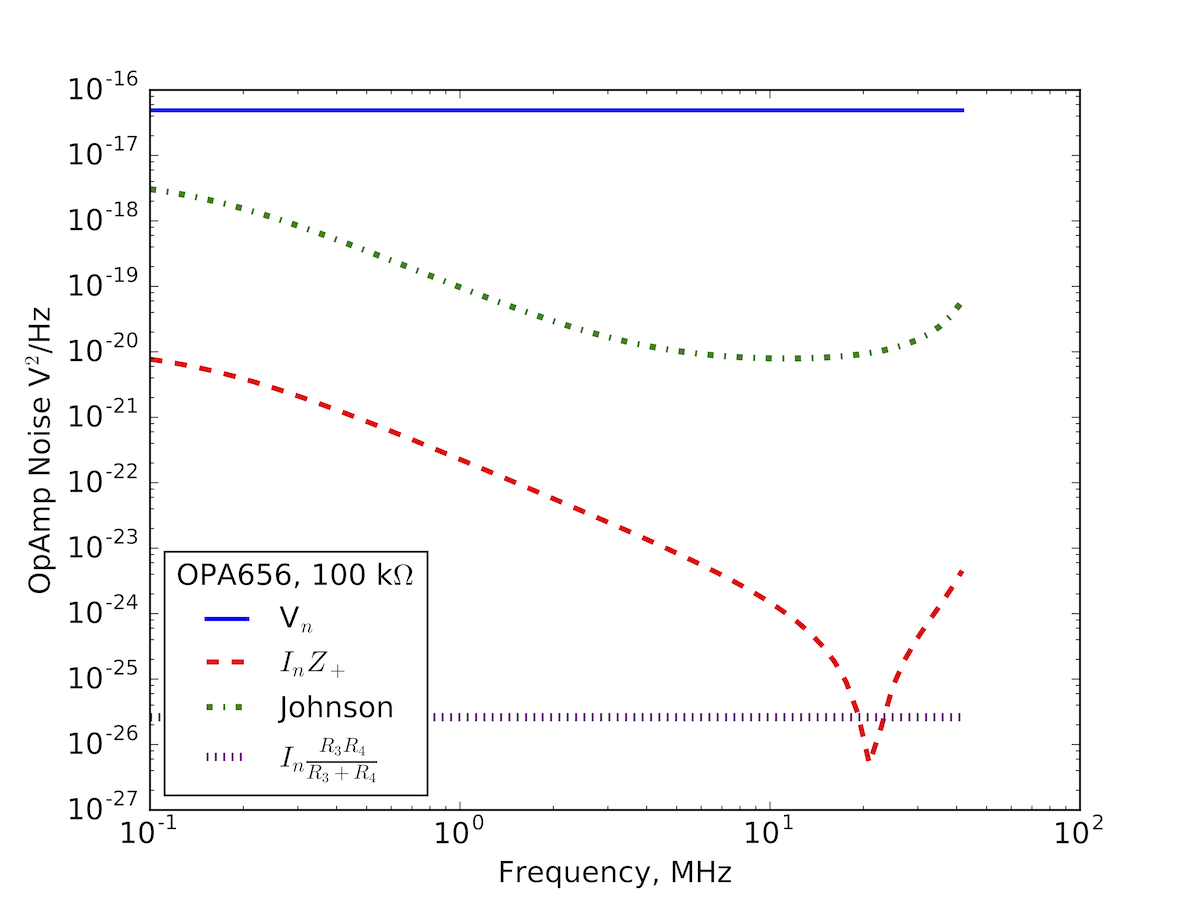}
		}
  \caption{The operational amplifier noise assuming different input impedance values. The noise is dominated by the voltage term regardless of the impedance used in the range considered. These simulations are for the 5-meter antenna only.}
 \label{OpAmpNoise:fig}
\end{figure}

Based on the Analog Devices Tutorial MT-049, adapted to the notation on our circuit diagram, the noise model is calculated according to:
\begin{equation}
V_{L,amp}^2=V_n^2+I_n^2 Z_+^2+I_n^2 \left(\frac{R_3 R_4}{R_3+R_4}\right)+4kT_0\left(Re(Z_+ )+R_3\left(\frac{R_4}{R_3+R_4}\right)^2+R_4\left(\frac{R_3}{R_3+R_4}\right)^2 \right),
\end{equation}

where $Z_+$ is the lump impedance seen from the terminal labeled ``+'', including the antenna impedance and $T_0$ is the physical temperature of the resistors, taken to be room temperature $T_0=$290 Kelvin. The last term in curly brackets is the Johnson noise of the system. The different noise terms, assuming $R_8=$12.8, 50 and 100k$\Omega$, are shown below. The different noise contribution, for a 5-meter antenna, are shown in Figure~\ref{OpAmpNoise:fig}. In all cases $Vn$ is, by far, the dominant source of noise.

\begin{figure}[htb]
\includegraphics[width=0.8\textwidth]{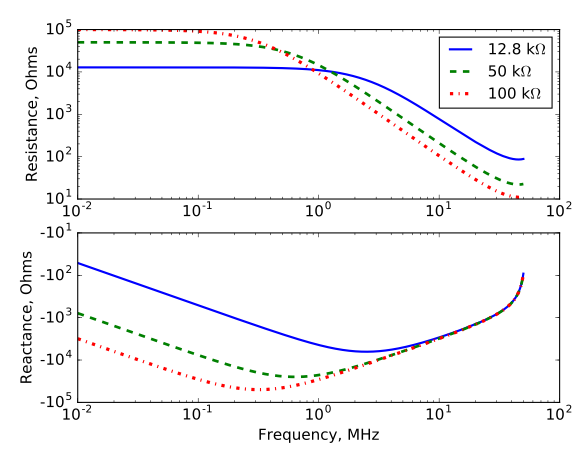}
\caption{Simulations of the impedance seen by the antenna terminals looking into the front-end design in Figure~\ref{OpAmpFrontEnd:fig} for R8 equal to 12.8, 50, and 100 k$\Omega$. }
\label{OpAmpImpedance:fig}
\end{figure}

Simulations were run to obtain the impedance as a function of frequency for 12.8 k$\Omega$, 50 k$\Omega$, and 100 k$\Omega$, simply by changing R8 in the design. The impedance of the amplifier as seen by the terminals of the antenna, is shown in Figure~\ref{OpAmpImpedance:fig}. 

\begin{figure}[htb]
\includegraphics[width=0.6\textwidth]{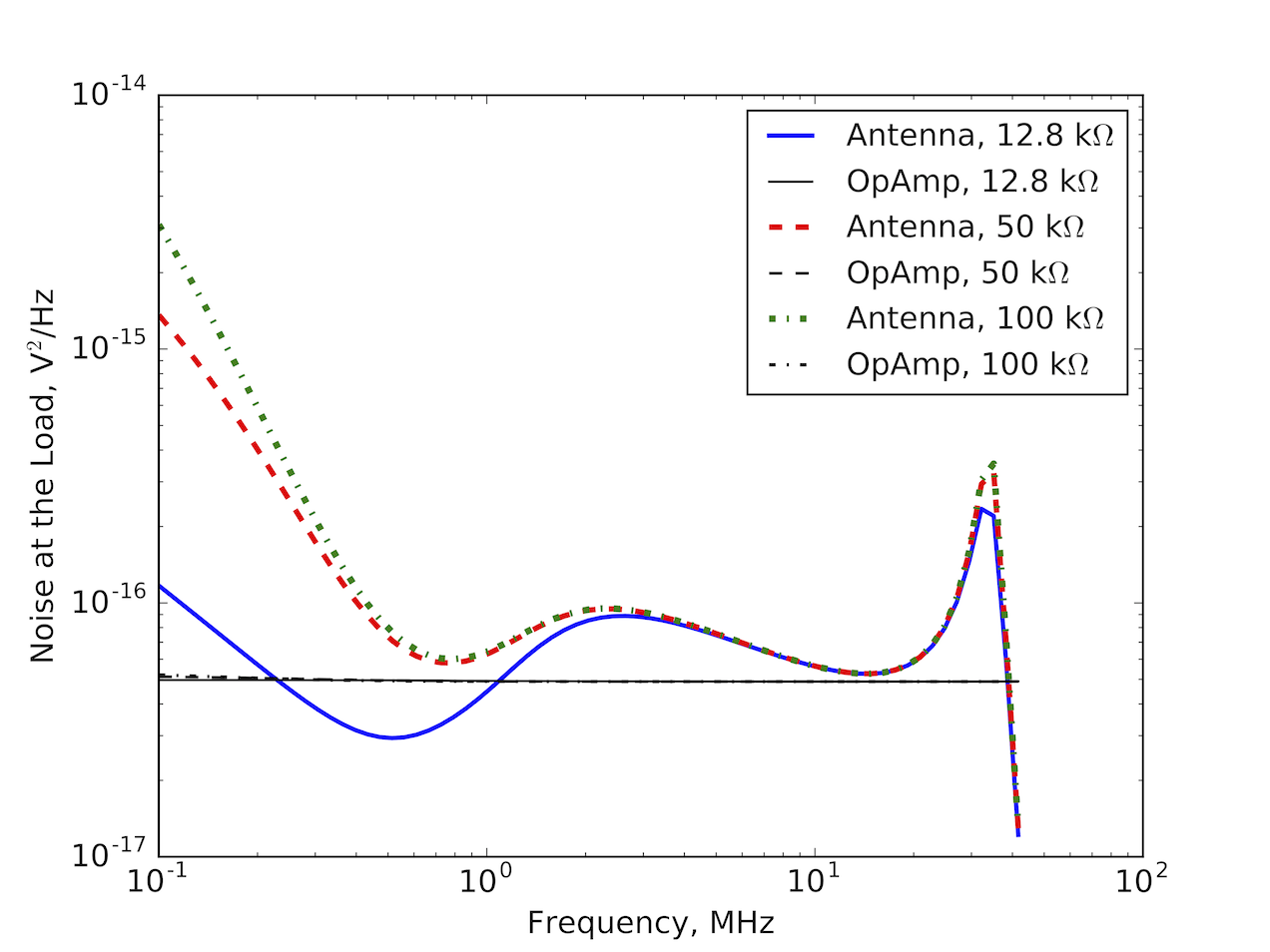}
\caption{Results for the noise due to the antenna and the amplifier noise for the front-end design in Figure~\ref{OpAmpFrontEnd:fig} assuming values of R8 equal to 12.8, 50, 100 k$\Omega$. }
\label{OPA656_results:fig}
\end{figure}

This OpAmp design has an impedance that is high at low frequencies and lower at high frequencies, which is the general desired direction for keeping the antenna noise above the amplifier noise across the band.  The voltage at the load is given by:
\begin{equation}
<v_L^2> =<v_A^2>\frac{|Z_L |^2}{|Z_L+Z_A |^2},
\end{equation}
where $Z_L$ is the impedance shown in Figure~\ref{OpAmpImpedance:fig}. The results for the antenna noise compared to the amplifier noise for the impedances considered here are shown in Figure~\ref{OPA656_results:fig}.

The results of the OpAmp approach indicate that the OpAmp has a higher antenna to amplifier noise ratio. The reason points to the impedance vs frequency of the OpAmp, which has a profile with high impedance at low frequencies and low impedance at higher frequencies. This is the general behavior needed for the antenna noise to dominate across the band. We also note that using an OPA656 impedance with R8 resistor value $>$~50~k$\Omega$ meets the desired design objectives laid out in the introduction. The antenna noise is above the amplifier noise with smooth behavior across the band that is not highly variable. The expected noise levels below 1 MHz can be as $>$~10~times higher than parts of the band above 1 MHz. However, frequencies below 1 MHz account for $<$~10\% of the full band.

\end{document}